\documentclass{aa} 
 
\usepackage{natbib} 
\usepackage[usenames]{color}
\bibpunct{(}{)}{;}{a}{}{,} 
\usepackage{amsmath} 
\usepackage{graphicx,txfonts} 
\usepackage{longtable} 
\usepackage{url} 
\usepackage{xspace}

\begin{document} 
\title{CoRoT's view on variable B8/9 stars: spots versus pulsations\thanks{The CoRoT space mission was developed and is operated by the French space agency CNES, with participation of ESA's RSSD and Science Programmes, Austria, Belgium, Brazil, Germany, and Spain.}$^,$\thanks{Based on observations made with the ESO telescopes at La Silla Observatory under the ESO Large Programme LP182.D-0356}$^,$\thanks{Based on observations made with the Mercator Telescope, operated on the island of La Palma by the Flemish Community, at the Spanish Observatorio del Roque de los Muchachos of the Instituto de Astrof\'isica de Canarias.}} 
\subtitle{Evidence for differential rotation in HD\,174648}
 
\author{P.~Degroote\inst{\ref{inst:leuven}} 
\and B.~Acke\inst{\ref{inst:leuven}}\thanks{Postdoctoral Fellow of the Fund for Scientific 
Research, Flanders} 
\and R.~Samadi\inst{\ref{inst:lesia}}
\and C.~Aerts\inst{\ref{inst:leuven},\ref{inst:nijmegen}} 
\and D.~W.~Kurtz\inst{\ref{inst:lancashire}} 
\and A.~Noels\inst{\ref{inst:liege}} 
\and A.~Miglio\inst{\ref{inst:liege},\ref{inst:birming}} 
\and J.~Montalb{\'a}n\inst{\ref{inst:liege}} 
\and S.~Bloemen\inst{\ref{inst:leuven}} 
\and A.~Baglin\inst{\ref{inst:lesia}} 
\and F.~Baudin\inst{\ref{inst:ias}}
\and C.~Catala\inst{\ref{inst:lesia}} 
\and E.~Michel\inst{\ref{inst:lesia}}
\and M.~Auvergne\inst{\ref{inst:lesia}}
} 
 
\institute{Instituut voor Sterrenkunde, K.U.Leuven, Celestijnenlaan 200D, B-3001 
Leuven, Belgium\label{inst:leuven} 
\and LESIA, Observatoire de Paris, CNRS UMR 8109, Universit\'e Pierre et Marie Curie, Universit\'e Denis Diderot, 5 place J. Janssen, 92105 Meudon, France\label{inst:lesia} 
\and Department of Astrophysics, IMAPP, University of Nijmegen, PO Box 9010, 6500 GL Nijmegen, The Netherlands\label{inst:nijmegen} 
\and Jeremiah Horrocks Institute of Astrophysics, University of Central Lancashire, Preston PR1 2HE, UK\label{inst:lancashire} 
\and Institut d'Astrophysique et de G\'eophysique Universit\'e de Li\`{e}ge, 
All\'{e}e du 6 Ao\^{u}t 17, B-4000 Li\`{e}ge, Belgium\label{inst:liege} 
\and School of Physics and Astronomy, University of Birmingham, Edgbaston, Birmingham B15 2TT, United Kingdom \label{inst:birming}
\and Institut d'Astrophysique Spatiale, CNRS/Universit\'e Paris XI UMR 8617, F-091405 Orsay, France\label{inst:ias}} 
 
\date{Received 1 March 2011; accepted 3 October 2011} 
\authorrunning{Degroote et al.} 
\titlerunning{CoRoT's view on variable B8/9 stars} 
 
\abstract 
{There exist few variability studies of stars in the region in the Hertzsprung-Russell diagram between the A and B-star pulsational instability strips. With the aid of the high precision continuous measurements of the CoRoT space satellite, low amplitudes are more easily detected, making a study of this neglected region worthwhile. } 
{We collected a small sample of B stars observed by CoRoT to determine the origin of the different types of variability observed.} 
{We combine literature photometry and spectroscopy to measure the fundamental parameters of the stars in the sample, and compare asteroseismic modelling of the light curves with (differentially rotating) spotted star models.} 
{We found strong evidence for the existence of spots and differential rotation in HD\,174648, and formulated hypotheses for their origin. We show that the distinction between pulsations and rotational modulation is difficult to make solely based on the light curve, especially in slowly rotating stars.} 
{} 

\keywords{Stars: oscillations; Stars: variables: early-type; Stars: fundamental 
parameters -- Stars: individual: HD\,174648} 
\maketitle 
 
\section{Introduction} 

The cool boundary of the instability strip of slowly pulsating B stars \citep[SPBs hereafter, ][]{waelkens1991} is not well constrained observationally. Excitation calculations predict that it should lie around $10\,000-11\,000$\,K \citep{dziembowski1993,miglio2007}, but this is highly dependent on parameters such as metallicity, core overshooting and rotation. An observational determination of the cool border can help to tune the input physics of the evolutionary models. It is, however, difficult to constrain the exact location of the instability strip purely based on observations, partially because the relevant fundamental parameters need to be known to an accuracy that can hardly be reached, but also because of the required observational effort to collect large enough samples of high precision photometry and/or time-resolved high-resolution spectroscopy to detect the low-amplitude gravity modes. In early attempts to map the variability in this area of the Hertzsprung-Russell diagram empirically, i.e., prior to the knowledge of the pulsation excitation mechanism, \citet{baade1989a,baade1989b} performed a search for line-profile variability in 22 dwarfs and giants of spectral types B8-B9.5. For \emph{none} of these targets, periodic line-profile variations were detected. Given the relatively high fraction of late B stars -- compared to earlier types -- the many thousands of stars observed in the CoRoT \citep{baglin2006} exoplanet fields and in the Kepler \citep{kepler} field are very promising to confront with the spectroscopic null results of \citet{baade1989a,baade1989b}. However, the effort in gathering a homogeneous set of fundamental parameters to select the late-B stars among those CoRoT and Kepler targets is an immense undertaking. 

Another reason to focus on late B stars is to study the co-existence of SPBs with Bp stars, HgMn stars and rapidly rotating stars \citep{briquet2007} to shed light on the underlying origins of their differences. Although a few bright SPBs have been observed with the MOST satellite \citep[e.g.,][]{aerts2006,gruber2009}, high-precision space-based observations of late  B  stars with well determined fundamental parameters are rare. In this respect, we explore the light curves of the late B stars measured in the CoRoT asteroseismology channel.

The structure of this paper is as follows. In Sect.\,\ref{sec:sample_description}, we present the properties of the individual stars in the sample. In Sect.\,\ref{sec:HD174648}, we focus on HD\,174648 and argue for the existence of spots on its surface. In the next section, we compare the light curve morphology of HD\,174648 to that of the other stars in the sample. We end with a discussion on the origins of the differences between them.
 
\begin{table*}\footnotesize 
\centering
\caption{CoRoT cool B star sample description for the asteroseismology programme of the mission.}\label{tbl:sample} 
\begin{tabular}{rlll|llccccl}\hline\hline 
\multicolumn{1}{c}{HD}& 
\multicolumn{1}{c}{field} &  
\multicolumn{1}{c}{SpT}\tablefootmark{a}   & 
\multicolumn{1}{c}{$m_V$\tablefootmark{a}} &      
\multicolumn{1}{c}{$E(B-V)$} & 
\multicolumn{1}{c}{$T_\mathrm{eff}$}& 
\multicolumn{1}{c}{$\log g$} & 
\multicolumn{1}{c}{$v_\mathrm{eq}\sin i$} &  
\multicolumn{1}{c}{$\pi$} &  
\multicolumn{1}{c}{$\theta$\tablefootmark{b}} & \multicolumn{1}{c}{Reference}\\ 
  & 
  & 
  & 
\multicolumn{1}{c}{(mag)} & 
\multicolumn{1}{c}{(mag)} & 
\multicolumn{1}{c}{(K)} & 
\multicolumn{1}{c}{(dex)} & 
\multicolumn{1}{c}{(km\,s$^{-1}$)} &  
\multicolumn{1}{c}{(mas)} & 
\multicolumn{1}{c}{(mas)} & 
\\\hline 
46179\tablefootmark{c} & SRa01 & B9V   & 6.69 & $0.02\pm0.01$          & $10900\pm500$                    & $4.0\pm0.2$         & $152\pm13$&               &                           & \citet{niemczura2009}\\             
                       &       &       &      & $0.02^{+0.04}_{-0.02}$ & $10730^{+600}_{-420}$            & $4.1^{+0.7}_{-0.8}$ &           & $3.58\pm0.78$ & $0.138^{+0.005}_{-0.004}$ & from SED\\                          
49677\tablefootmark{c} & SRa01 & B9    & 8.06 & $0.03^{+0.02}_{-0.02}$ & \phantom{1}$9185^{+270}_{-200}$  & $4.0^{+0.3}_{-0.4}$ & -         & -             & $0.088^{+0.001}_{-0.001}$ & from SED\\                          
181440                 & LRc01 & B9III & 5.48 & $0.02\pm0.01$          & $11200\pm500$                    & $3.5\pm0.2$         & $58\pm1$  &               &                           & \citet{niemczura2009}\\             
                       &       &       &      &                        & $11500\pm500$                    & $3.6\pm0.5$         & $55\pm7$  &               &                           & \citet{lefever2010}\\               
                       &       &       &      & $0.06^{+0.04}_{-0.05}$ & $11060^{+600}_{-660}$            & $4.0^{+0.8}_{-0.9}$ &           & $7.45\pm0.28$ & $0.252^{+0.010}_{-0.009}$ & from SED\\                          
182198                 & LRc01 & B9V   & 7.94 & $0.04\pm0.01$          & $11450\pm500$                    & $3.5\pm0.2$         & $25\pm1$  &               &                           & \citet{niemczura2009}\\             
                       &       &       &      &                        & $11000\pm500$                    & $3.1\pm0.1$         & $23\pm1$  &               &                           & \citet{lefever2010}\\               
                       &       &       &      & $0.03^{+0.02}_{-0.02}$ & $11290^{+410}_{-500}$            & $3.6^{+0.8}_{-1.1}$ &           & $0.89\pm0.74$ & $0.109^{+0.003}_{-0.003}$ & from SED\\                          
170935                 & LRc02 & B8    & 7.38 & $0.20\pm0.01$          & $10300\pm500$                    & $3.0\pm0.2$         & $271\pm16$&               &                           & \citet{niemczura2009}\\             
                       &       &       &      & $0.20^{+0.03}_{-0.02}$ & \phantom{1}$9750^{+600}_{-290}$  & $3.5^{+0.6}_{-1.1}$ &           & $2.77\pm0.58$ & $0.145^{+0.002}_{-0.003}$ & from SED\\                          
174648                 & SRc02 & B9    & 8.82 & $0.14^{+0.04}_{-0.06}$ & $10050^{+780}_{-740}$            & $4.0^{+0.9}_{-2.0}$ &           & $4.06\pm0.88$ & $0.066^{+0.002}_{-0.002}$ & from SED\\                          
                       &       &       &      & $0.20\pm0.04$          & $11000\pm1000$                   & $4.2\pm0.3$         & $288\pm6$ &               &                           & from HERMES spectra\\\hline\hline   
\end{tabular} 
 
\tablefoottext{a}{SIMBAD spectral type or visual magnitude} 
\tablefoottext{b}{angular diameter} 
\tablefoottext{c}{binary} 

\end{table*}

\begin{figure}
 \includegraphics[width=\columnwidth]{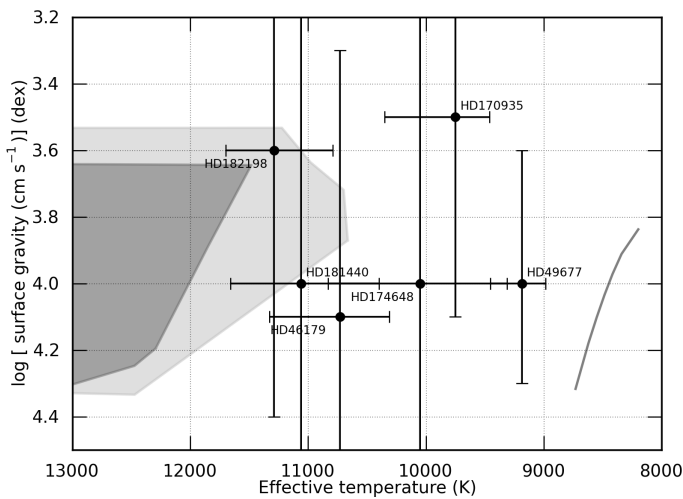}
\caption{Location of the sample in the $T_\mathrm{eff}-\log g$ diagram as determined with multicolour photometry. The right grey line represents the blue edge of the $\delta$\,Sct instability strip. The light and dark grey regions represent SPB instability strips with $Z=0.01$ (dark grey) and $Z=0.02$ (light grey) and were taken from  \citet{miglio2007a}.}
\label{fig:hrdiagram}
\end{figure}
 
\section{Sample description}\label{sec:sample_description} 

The stars in the CoRoT late B star sample, being secondary asteroseismology targets, are not well known. For four of the stars the spectral subclasses are not listed in the literature. The basic parameters of the complete sample so far is presented in Table\,\ref{tbl:sample} and Fig.\,\ref{fig:hrdiagram}, and part of the chemical compositions of stars with known abundances are provided in Table\,\ref{tbl:abundances}. In the framework of a preparatory programme for CoRoT \citep{solano2005}, the objects HD\,46179, HD\,181440, HD\,182198 and HD\,170935 were analysed spectroscopically by \citet{niemczura2009} and \citet{lefever2010}. To obtain a consistent determination of the fundamental parameters ($T_\mathrm{eff}, \log g$, and $E(B-V)$) for the whole sample, we fitted Kurucz atmosphere models \citep{kurucz1993} to the literature photometry for these stars, assuming the average extinction curve for the Milky Way with $R_v=3.1$ \citep{chiar2006} and solar metallicity. We used the catalogues of \citet{cat_tycho}, \citet{cat_2mass}, \citet{cat_usnob1}, \citet{cat_uvby}, \citet{cat_gcpd}, \citet{cat_td1}, \citet{cat_pasp}, \citet{cat_tass}, and \citet{cat_eso}, and calibrated the photometry using the Vega model of \citet{bohlin2004} and the zero points of \citet{maizapellaniz2007}. The best model and 90\% confidence intervals for the parameters were determined via an interpolation of the atmosphere grid in the three parameters $T_\mathrm{eff}$, $\log g$ and $E(B-V)$. The Introducing a scaling factor representing the angular diameter of the star, a $\chi^2$ statistic was used to determine the goodness-of-fit of $\sim$ 1 million models randomly chosen within the grid boundaries. Using the same statistic and all explored models,  90\% confidence intervals on the parameters could be derived. We checked these results with a Nelder-Mead minimization routine and Monte-Carlo simulations. Given that the inclusion of UV fluxes does not always result in a better determination of the fundamental parameters, which is partly because the systematic errors are larger, but also because of poor agreement between observed and predicted colours from atmosphere models, we did not include them.
When possible, the estimated parameters were cross-correlated with results obtained via different methods and found in the literature: the effective temperatures were also determined via Str\"omgren multicolour photometry using the calibration scheme of \citet{moon1985}; the calibration of \citet{cramer1984} was applied to Geneva photometry. Where spectra were available the reddening was estimated using the diffuse interstellar bands (DIBs) at 6196\,\AA, 6613\,\AA\ and 5780\,\AA\ using the calibration of \citet{luna2008}. We compared these values to the reddening found via the spectral energy distribution (SED) fit, to ensure that the effective temperature determinations are meaningful. For all stars, we found consistent sets of parameters, but, given the unknown systematic uncertainties, we also list the literature values besides the one we obtained from SED fitting.
 
\begin{table*}\footnotesize 
\centering\caption{Element abundances for the stars in the CoRoT cool B star sample.}\label{tbl:abundances} 
\begin{tabular}{lllllllllllllllll}\hline\hline 
Star     & He   &  C     &  O    &  Mg   & Al  &  Si   & P   &  S     &  Ca    & Ti    & V    &  Cr    & Fe    & Ni   & Ba    \\ \hline 
HD\,170935 &10.79 &        &       &  8.18 &     &  7.36 &     &        &        & 4.64  &      &  5.66  & 7.57  &      &       \\ 
HD\,181440 &10.90 &        &       &  7.80 &     &  7.24 &     &        &        & 4.36  &      &  5.09  & 6.85  &      &       \\ 
HD\,182198 &10.93 &  8.38  &  7.61 &  7.48 & 6.56&  7.34 & 5.28&  7.06  &  6.15  & 4.61  & 4.55 &  5.65  & 7.27  & 6.45 & 2.56  \\ 
HD\,46179  &11.04 &        &  8.85 &  7.58 &     &  7.60 &     &        &        & 4.61  &      &  5.79  & 7.29  &      &       \\\hline 
Sun      &10.93 &  8.39  &  8.66 &  7.53 & 6.37&  7.51 & 5.36&  7.14  &  6.31  & 4.90  & 4.00 &  5.64  & 7.45  & 6.23 & 2.17  \\\hline\hline 
\end{tabular} 
\tablefoot{The abundance of an element $X$ is given as $\log\epsilon(X)$, where $\log\epsilon(\mathrm{H})=12$ \citep[taken from ][]{niemczura2009}. Solar abundances from \citet{asplund2005}.}
\end{table*}

As a consequence of CoRoT's target selection procedure, the sample of stars is fairly unbiased, with the only common property that the stars are late  B  and are located near the galactic plane. At least three binaries are included, which is expected from the high binarity fraction of massive stars \citep[e.g., ][]{preibisch2001}. Among the targets, there are four slow rotators, one moderate rotator and three fast rotators. The different surface gravities suggest that the targets are spread over the main sequence. 
 
In the following, we focus on HD\,174648, which is a fast rotator showing clear variability with a simple frequency spectrum, which is interpreted as due to spots. We then compare the results with the light curve analysis of the other (pulsating) stars.
 
\section{HD\,174648} \label{sec:HD174648}
 
\subsection{Fundamental parameters}\label{sect:coolbstars:fundamental_parameters} 

Two spectra are available from the fibre-fed \'echelle spectrograph HERMES \citep{hermes} on the 1.2-m Mercator telescope (La Palma). The spectrograph covers a range from 3800\,\AA\ to 8800\,\AA\ with a resolution of $\sim$$80\,000$. The spectra were measured on 2010 April 18 with a S/N of $\sim$100 and $\sim$150. A first estimate of the rotation velocity was obtained using the Fourier method \citep{simondiaz2007} applied to the profile of the Mg\,\textsc{ii}\,4481\AA\ line. We estimate the uncertainties via Monte Carlo simulations in the following way. In 10\,000 data sets we add random Gaussian noise and vary the wavelength range on the left and right of the line's central wavelength according to a uniform distribution defined on the interval [-1,+1]\,\AA\ with respect to the interval chosen by eye. From the first zero of the Fourier transform, we obtain $v_\mathrm{eq}\sin i = 288\pm6$\,km\,s$^{-1}$; the velocity connected to the second zero of the Fourier transform is $v_2\sin i = 156\pm3$\,km\,s$^{-1}$. \citet{reiners2002} derived a relation between the ratio $q_2/q_1$ of the two first zeros in the Fourier transform and the differential rotation rate. For slowly and rigidly rotating massive stars, and assuming a linear limb darkening law, this ratio is approximately 1.76, which is on the low side of our estimate $q_2/q_1=1.84\pm0.05$. This suggests the presence of differential rotation with polar acceleration. However, the Mg\,\textsc{ii} line is actually a narrow blend, typically resulting in an overestimation of the $v_\mathrm{eq}\sin i$ by $~10$\,km\,s$^{-1}$ at these rotation velocities, and raising the $q_2/q_1$ value appreciably. We checked if we could improve the shape of the rotational kernel by taking more lines into account \citep[e.g., as was done in ][]{reiners2004b}, but this was not case: there are far too few lines, and the continuum estimation for these few lines is not reliable enough.
 
\begin{figure*} 
\centering\includegraphics[width=0.98\columnwidth]{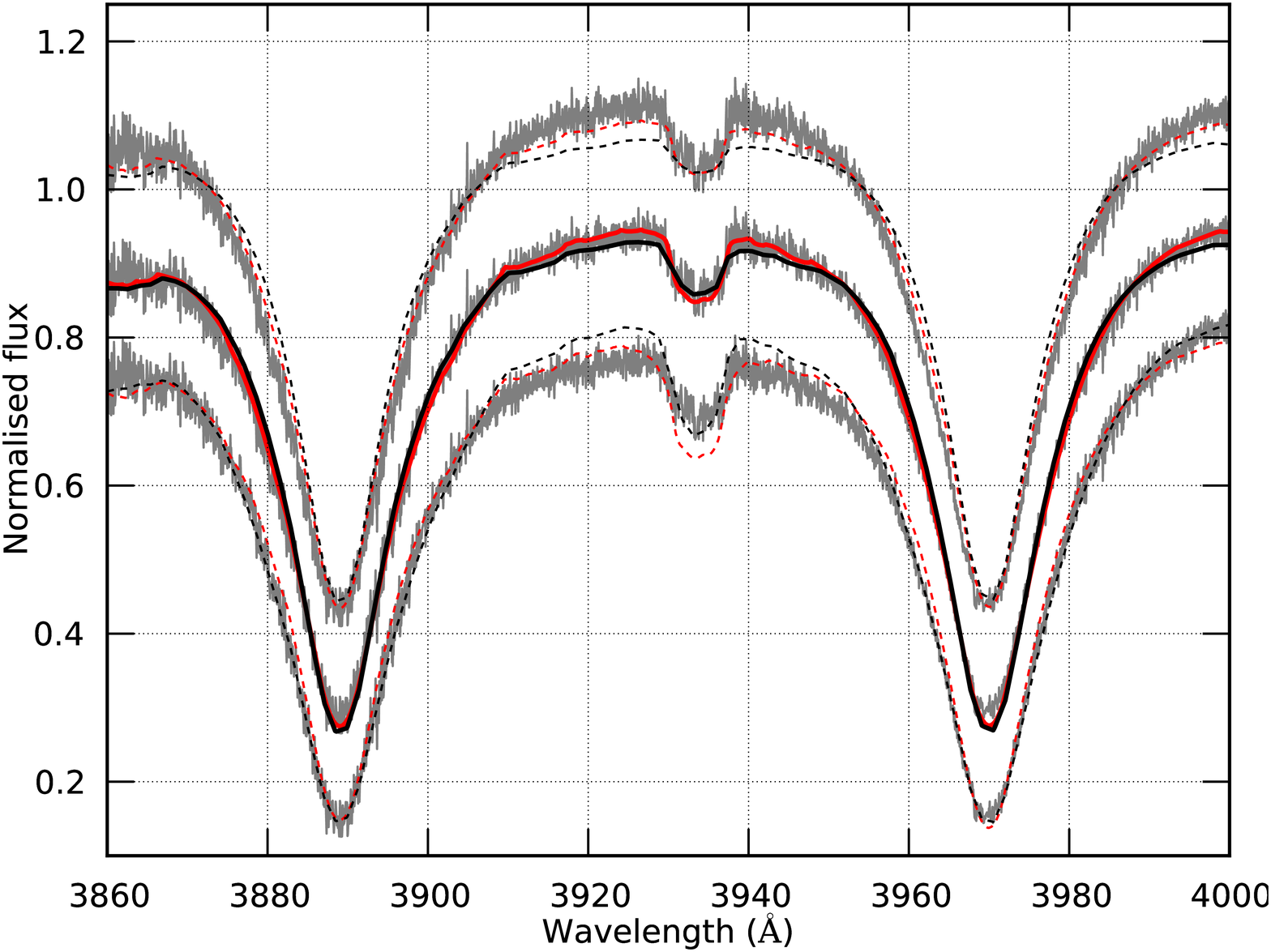} 
\centering\includegraphics[width=0.98\columnwidth]{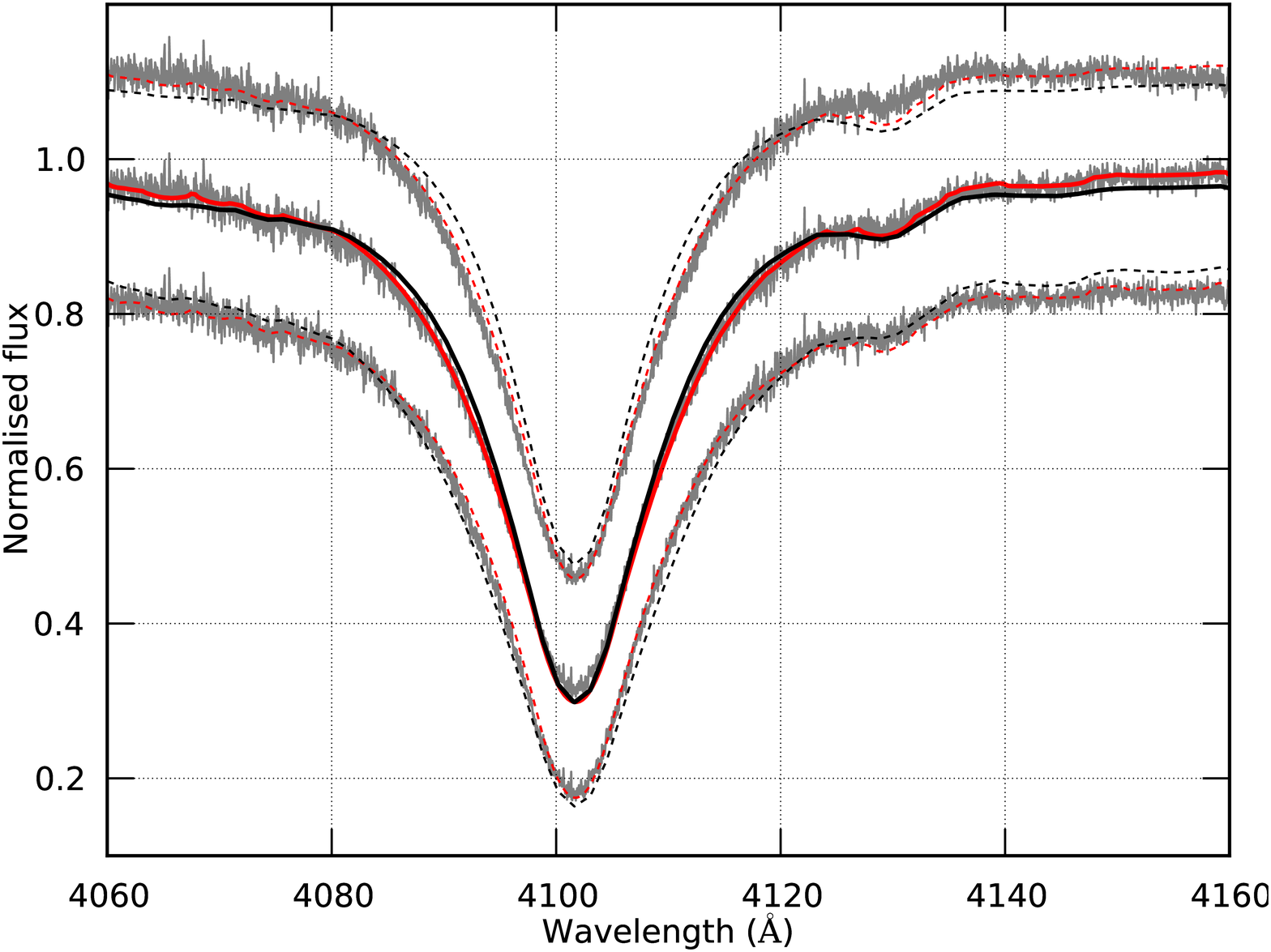} 

\centering\includegraphics[width=0.98\columnwidth]{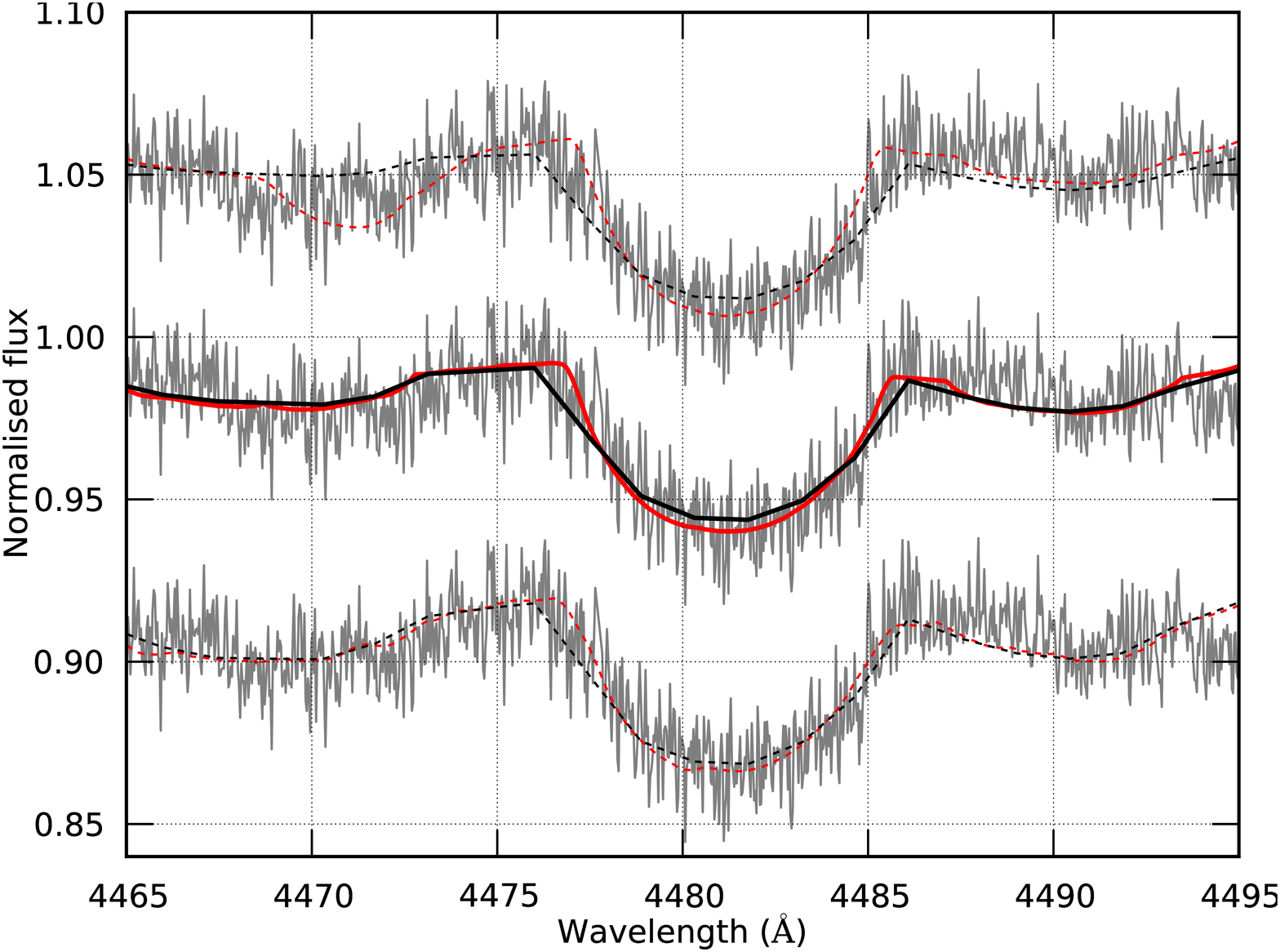} 
\centering\includegraphics[width=0.98\columnwidth]{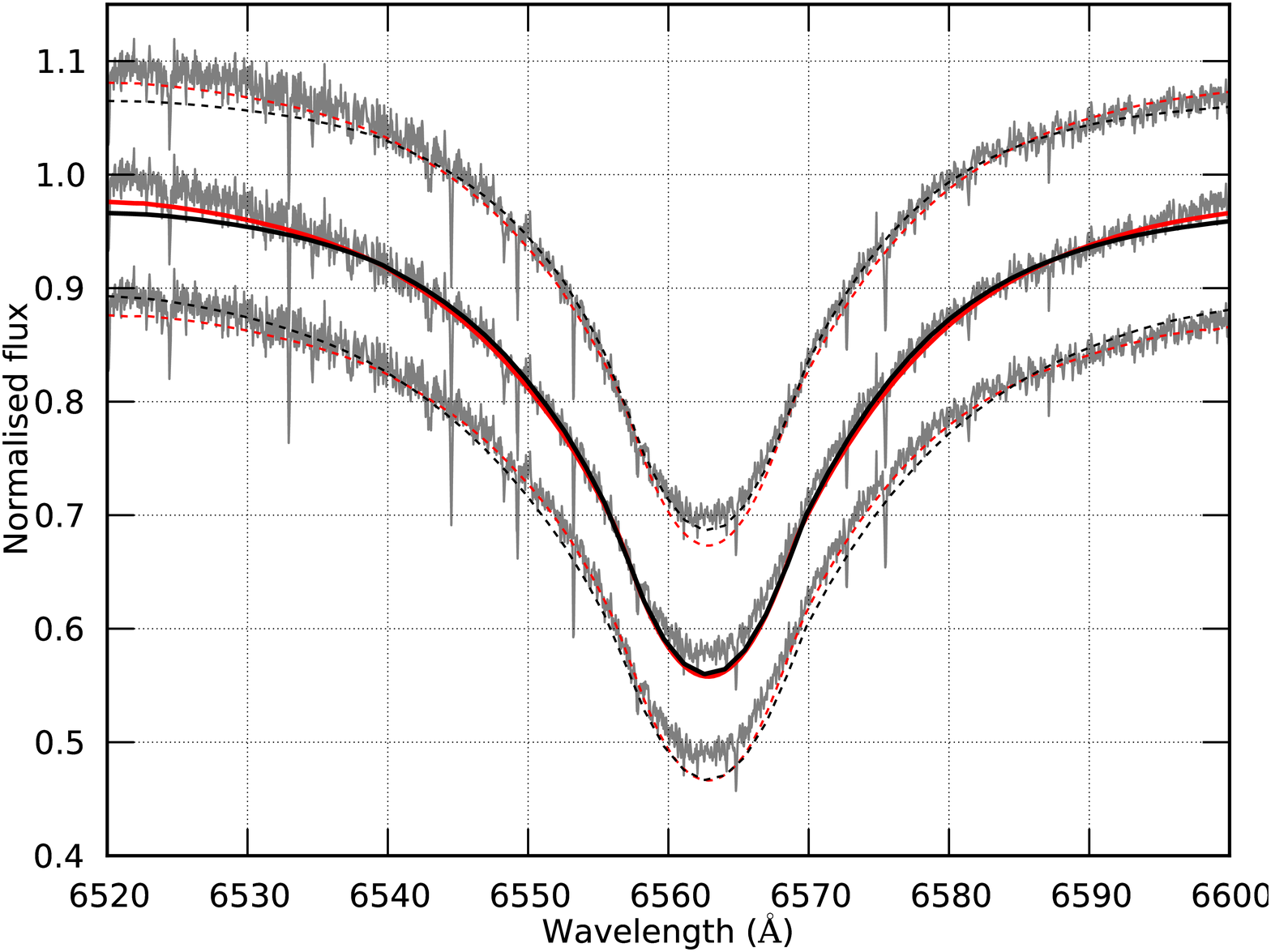} 
\caption{Comparison between model template spectra (black), synthetic spectra (red) additionally deformed according to model parameters resulting from the spot model fit (see text), and the observed profiles (grey) (from top to bottom and left to right: Ca\,\textsc{ii}\,3933\AA\ with H$\zeta$ and H$\epsilon$, H$\delta$, Mg\,\textsc{ii}\,4481\AA\ and H$\alpha$). In each panel, the top black spectrum shows the comparison to a template with parameters $T_\mathrm{eff}=11\,500\,K$ and $\log g=4.0$, the middle to the template $T_\mathrm{eff}=10\,500\,K$ and $\log g=4.0$ and the bottom to $T_\mathrm{eff}=9\,500\,K$ and $\log g=4.0$. The top and bottom red dashed lines show models (with  $\log g=4.0$) compatible with the photometry with $T_\mathrm{eff}=11\,580$\,K and $T_\mathrm{eff}=10\,357$\,K, respectively.} 
\label{fig:coolbstars:hd174648:fundpars} 
\end{figure*}

To determine the effective temperature and surface gravity of the star, we compared the observed spectral lines with an ATLAS12 grid of synthetic spectra \citep{palacios2010}, after broadening them according to a $v_\mathrm{eq}\sin i=288$\,km\,s$^{-1}$ rotational profile and an instrumental profile with a full width at half maximum of 0.08\,\AA. We assumed a parabolic shape of the continuum, via a Nelder-Mead Simplex minimization method, where we minimized the residuals of the normalised observed spectral profile and the trial synthetic profile, via a $\chi^2$-like statistic, 
\[\chi'^2 = \frac{1}{N}\sum_{\lambda=\lambda_0}^{\lambda_n} \left(\frac{F_{\mathrm{obs}}(\lambda)}{a\lambda^2 + b\lambda +c} - F_{\mathrm{syn}}(\lambda)\right)^2.\] 
The high rotation rate smears out the large majority of the lines, making abundance determination impossible. The most prominent lines in the observed spectrum are the Balmer lines, the Mg\,\textsc{ii}\,4481\AA\ line and the Ca\,\textsc{ii}\,3933\AA\ line. The equivalent width of the latter is the most sensitive one to temperature, based on the synthetic spectra. Moreover, the low reddening minimizes interstellar contributions. Therefore, we perform an independent check to derive the temperature only using the equivalent width of this feature. We calculated the equivalent width of the observed line by integrating the feature between two points outside the line region. The equivalent width distribution is again recovered via Monte Carlo simulations. The 1$\sigma$ error on the equivalent width translates to effective temperatures between 10\,000\,K and 10\,750\,K. Taking 3$\sigma$ extends this interval to $9\,750-13\,750$\,K, but at $T_\mathrm{eff}\gtrsim11\,000$\,K, the Ca\,\textsc{ii} feature becomes less sensitive to the temperature. In Fig.\,\ref{fig:coolbstars:hd174648:fundpars} we show the fit to a collection of spectral lines, with synthetic spectra corresponding to 9500\,K, 10500\,K and 11500\,K. We conclude that different methods give a consistent estimate of $T_\mathrm{eff}\approx 10500$\,K, which is also consistent with the appearance of a weak HeI line at 4471\,$\AA$ and normal He abundance. Finally, from the width of the Balmer wings, we deduce a surface gravity of $\log g=4.2\pm0.3$.                                                                                                         
 
On the basis of the CoRoT light curve morphology \citep[see, e.g., ][]{strassmeier1992} and the star's location in the HR diagram, we investigated two processes that could possibly lie at the origin of cyclical modulation: pulsations and spots. 
 
\subsection{Nonlinear pulsations}

\begin{table*}[htb]
\caption{Frequency analysis of the CoRoT light curve of HD\,174648 after correcting for discontinuities and removal of the satellite's orbital signal.}\label{tbl:HD174648_freqs}
\centering \begin{tabular}{lrrrrl}
\hline\hline
\multicolumn{1}{c}{ID}     
& \multicolumn{1}{c}{$f$}      
& \multicolumn{1}{c}{$a$}   
& \multicolumn{1}{c}{$\phi$}     
& \multicolumn{1}{c}{S/N} 
& \multicolumn{1}{c}{notes}\\
& \multicolumn{1}{c}{d$^{-1}$} 
& \multicolumn{1}{c}{ppm}     
& \multicolumn{1}{c}{rad/2$\pi$} 
&  &\\\hline
$f_1$    & $  3.79144 \pm  0.00008$ &$1829.3 \pm 6.5$&  $-0.273\pm  0.003$&  27.42 &\\  
$f_2$    & $  3.84083 \pm  0.00007$ &$1356.7 \pm 3.8$&  $-0.313\pm  0.003$&  31.27 &\\  
$f_3$    & $  7.58282 \pm  0.00008$ &$ 627.0 \pm 2.2$&  $ 0.005\pm  0.003$&  31.66 & 2$f_1-0.00007$\\ 
$f_4$    & $  7.68346 \pm  0.00032$ &$ 156.0 \pm 2.1$&  $-0.050\pm  0.013$&  23.76 & 2$f_2-0.0018$\\  
$f_5$    & $  0.23672 \pm  0.00112$ &$  56.4 \pm 2.1$&  $ 0.296\pm  0.047$&   5.88 &\\  
$f_6$    & $ 11.37273 \pm  0.00116$ &$  41.5 \pm 2.1$&  $-0.029\pm  0.050$&  10.64 & 3$f_1-0.0016$\\  
$f_7$    & $  0.30554 \pm  0.00131$ &$  43.1 \pm 2.1$&  $-0.309\pm  0.055$&   5.29 &\\  
$f_8$    & $  3.95753 \pm  0.00153$ &$  33.0 \pm 2.1$&  $-0.003\pm  0.065$&   5.80 &\\  
$f_9$    & $  0.09974 \pm  0.00179$ &$  26.3 \pm 2.0$&  $-0.021\pm  0.075$&   4.09 &\\  
$f_{11}$ & $  7.63157 \pm  0.00205$ &$  25.0 \pm 2.0$&  $-0.055\pm  0.086$&   6.26 &$f_1$+$f_2-0.0007$\\  
$f_{12}$ & $  0.54299 \pm  0.00229$ &$  19.5 \pm 2.0$&  $ 0.192\pm  0.097$&   3.65 &$f_5$+$f_7-0.0007$\\  
$f_{16}$ & $  0.27052 \pm  0.00303$ &$  21.0 \pm 2.0$&  $ 0.110\pm  0.128$&   3.19 &$f_{13}-f_{15}-0.0003$\\
$f_{18}$ & $  6.02177 \pm  0.00345$ &$  14.4 \pm 2.0$&  $-0.786\pm  0.145$&   3.82 &$2f_{17}$\\  
$f_{19}$ & $ 12.81640 \pm  0.00354$ &$  13.5 \pm 2.0$&  $ 0.530\pm  0.149$&   4.13 &\\\hline\hline
 \end{tabular}                                                    
\tablefoot{The frequencies, amplitudes and phases were optimized via nonlinear fitting after every prewhitening stage, using only the last 5 identified frequencies.}
\end{table*}                                                      
                                                                  
\begin{figure}                                                    
\includegraphics[width=\columnwidth]{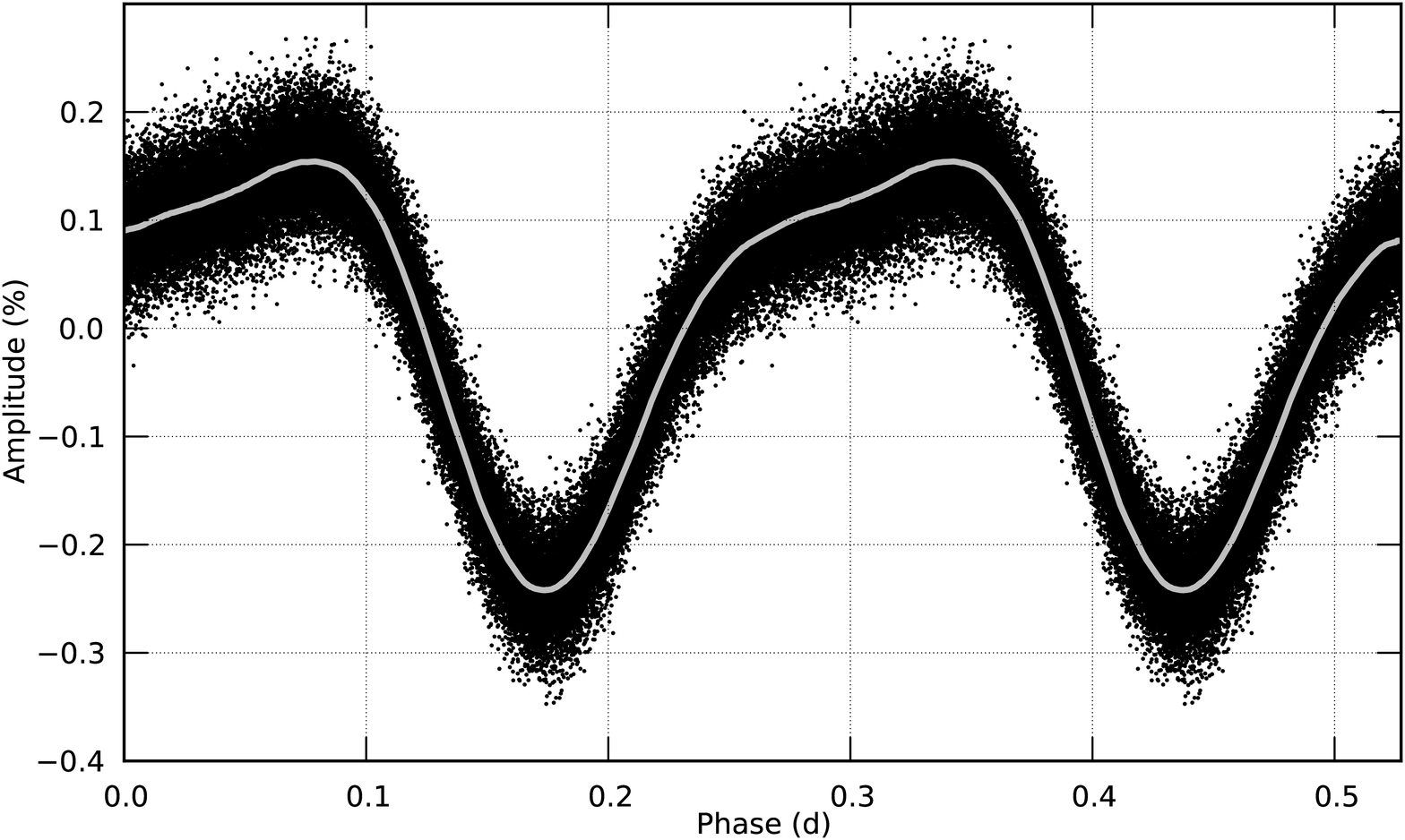}       
\includegraphics[width=\columnwidth]{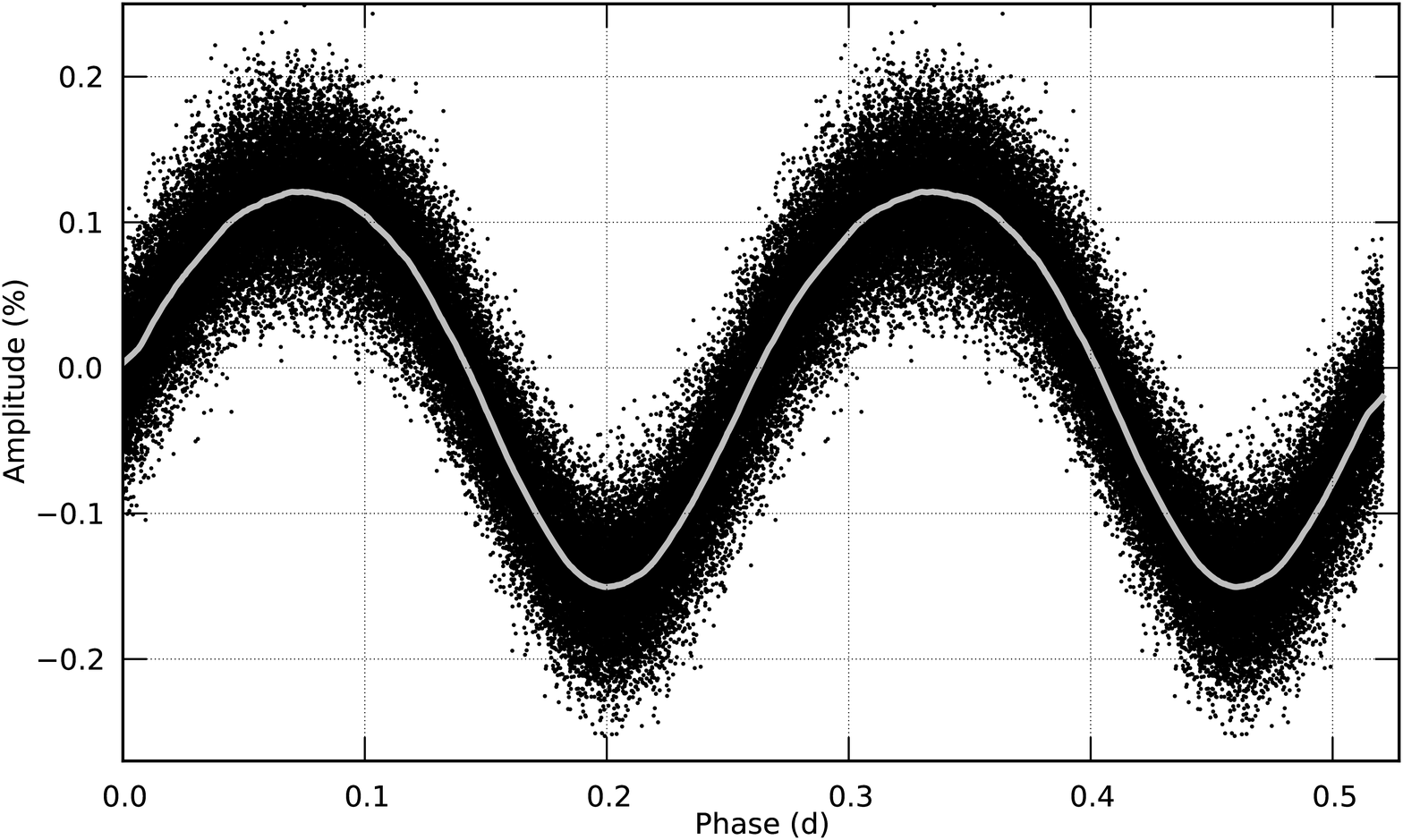}       
\caption{Diagrams of the CoRoT light curve of HD\,174648, folded on the dominant frequencies $f_1$ (top) and $f_2$ (bottom), after prewhitening all other detected frequencies. The top phase diagram clearly illustrates the non-sinusoidal shape of the variations for $f_1$; the nonsinusoidal shape for $f_2$ in the bottom diagram is less obvious, but is clearly demonstrated by the presence of the first harmonic of this frequency at $2f_2$, labelled $f_4$ in Table\,\ref{tbl:HD174648_freqs}. } 
\label{fig:coolbstars:HD174648:phasediagram}                      
\end{figure}

Before performing a full frequency analysis of the CoRoT light curve of HD\,174648, we corrected it for discontinuities, and minimized the influence of the satellite's orbital signal \citep[see, e.g., ][]{degroote2010b}. We performed a frequency analysis with nonlinear optimization after every prewhitening stage, using only the last 5 frequencies found at that stage, to account for the closely spaced peaks. Although the full frequency analysis resulted in the detection of 28 frequencies, we only list those with a S/N $>4$ or possible combinations in Table\,\ref{tbl:HD174648_freqs}. The first two frequencies $f_1$ and $f_2$ are closely spaced ($\Delta f\approx 1.2/T$), and two and one of their harmonics are recovered, respectively ($f_3 = 2f_1$, $f_4 = 2f_2$, and $f_6 =3f_1$). The phase diagrams of these frequencies, after prewhitening all other detections, are shown in Fig.\,\ref{fig:coolbstars:HD174648:phasediagram}. These, and the presence of the harmonics of these two frequencies found in the Fourier transforms, reveal unusual behaviour compared to the sinusoidal phase diagrams of confirmed SPBs \citep[see, e.g., ][]{decat2002}. We cannot make any firm conclusions on the two low and independent frequencies $f_5$ and $f_7$, because their period is long compared to the total time of observations, and they could thus be influenced by instrumental effects. 
                                                                  
Although SPB models have eigenfrequencies in the region of interest, the corresponding $\ell$-values of 5 and higher result in strong geometric cancellation, making high amplitudes of $\sim$0.1\% improbable. In fast rotators, geometric cancellation is even more effective than for slow rotators because of two additional effects: first, the modes are more concentrated around the equator \citep[e.g., ][]{reese2009}, and second, there is less flux coming from these regions due to gravity darkening \citep[e.g., 5 times less at $0.9v_\mathrm{crit}$, ][]{aerts2004b}. To explain the dominant frequencies with a beating phenomenon due to pulsations, it is thus necessary to shift low-degree model frequencies of slowly rotating SPB stars to higher frequencies; cf. the detection of three pulsation modes around 2.5\,d$^{-1}$ for the rapidly rotating SPB star HD\,121190 \citep[]{aerts2005}. 
                                                                  
Rotationally shifting the low-degree frequencies from $\sim$1\,d$^{-1}$ to $\sim$4\,d$^{-1}$ requires a high rotational angular velocity: in the case of an $\ell=1,m=1$ gravity mode, and only retaining first order effects to obtain an upper limit, we get 
\[\Omega = \frac{\Delta f}{m \beta_{\ell n}} \approx 6 \, \mathrm{d}^{-1}.\] 
The maximum possible angular break-up velocities for cool B  stars are reached at the ZAMS (when they are most compact), where values up to $\sim$3.5\,d$^{-1}$ can be expected. As a star evolves the angular break-up velocity decreases.  The required angular velocity to shift the frequencies can be lowered by increasing the mode degree, and more importantly the azimuthal order. At the same time, the observed frequency spectrum should lose all its structure, since the separation between the modes becomes higher. This is in sharp contrast to the observed frequency spectrum with the two closely spaced frequencies. 
                                                                  
We conclude that it is highly unlikely that the observed frequency spectrum of HD\,174648 is caused by stellar pulsation: the effective temperature is far too low to excite pressure modes in that region, and the frequency spectrum is too structured to originate from rotationally shifted nonradial gravity modes. We thus investigate the possibility that the observed variability does not arise from pulsations, but from rotational modulation. 

\subsection{Spots and differential rotation} 
 
\subsubsection{Observational constraints} 

The appearance of two high-amplitude, closely-spaced frequencies is a natural signature of differentially rotating spots \citep{lanza1993}, especially if their second harmonics are also prominent, but higher harmonics are not \citep{clarke2003}. The frequency splitting $\Delta f\approx0.05$\,d$^{-1}$ of HD\,174648 suggests that the periods of rotation at the spot latitudes differ by $\sim$1\%. 
 
\begin{figure*} 
\centering
\includegraphics[width=0.98\columnwidth]{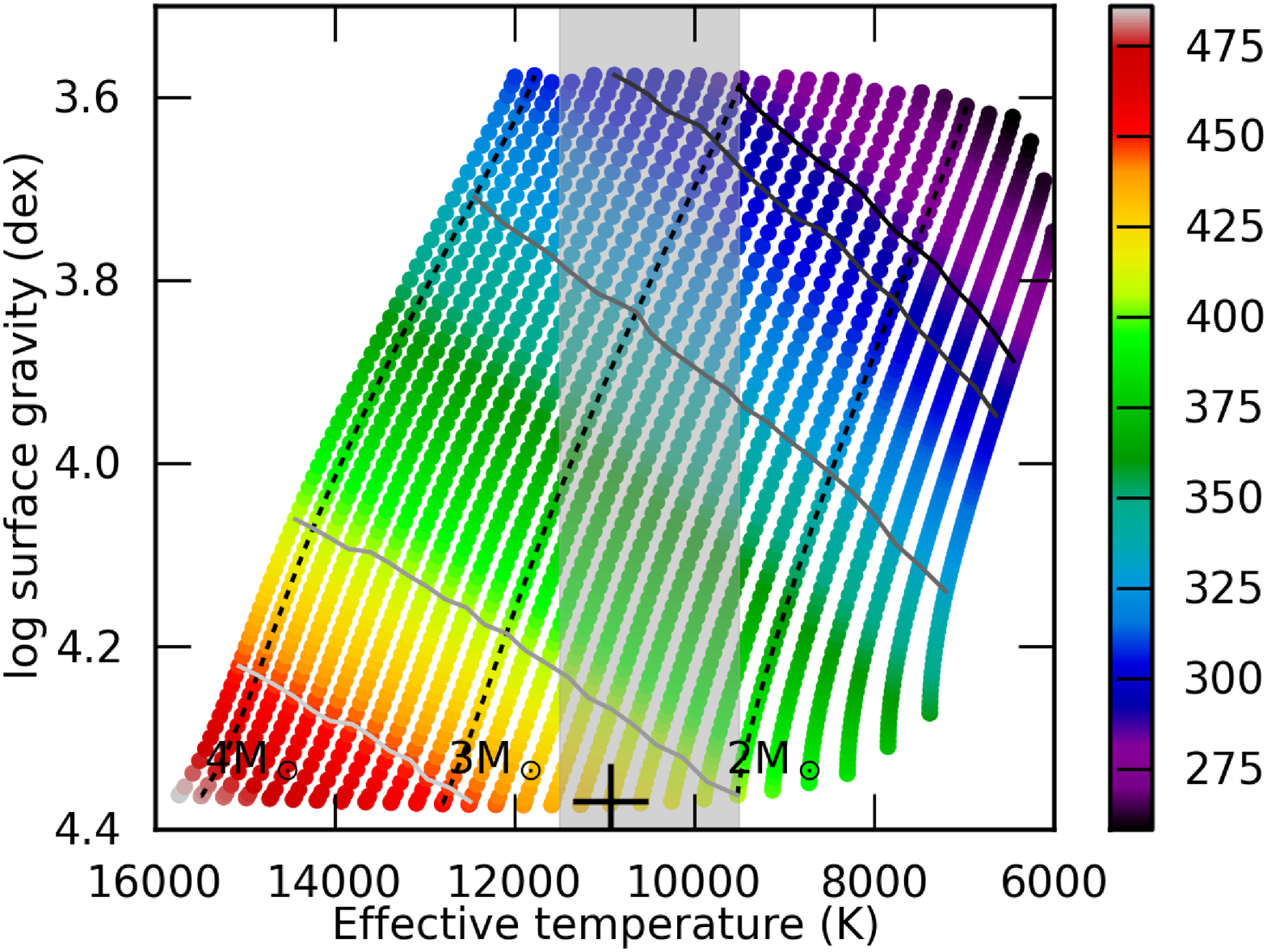} 
\includegraphics[width=0.98\columnwidth]{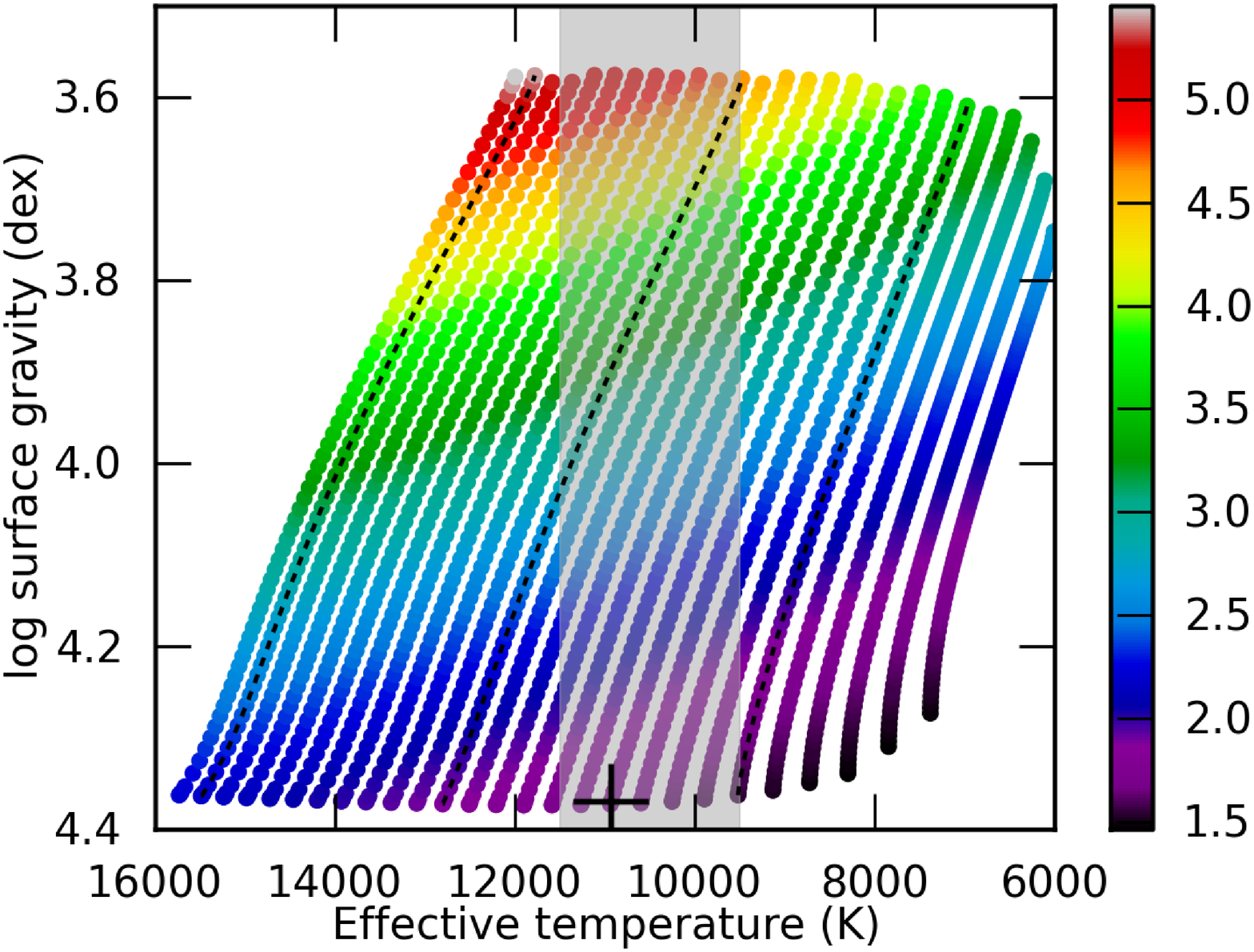} 
\includegraphics[width=0.98\columnwidth]{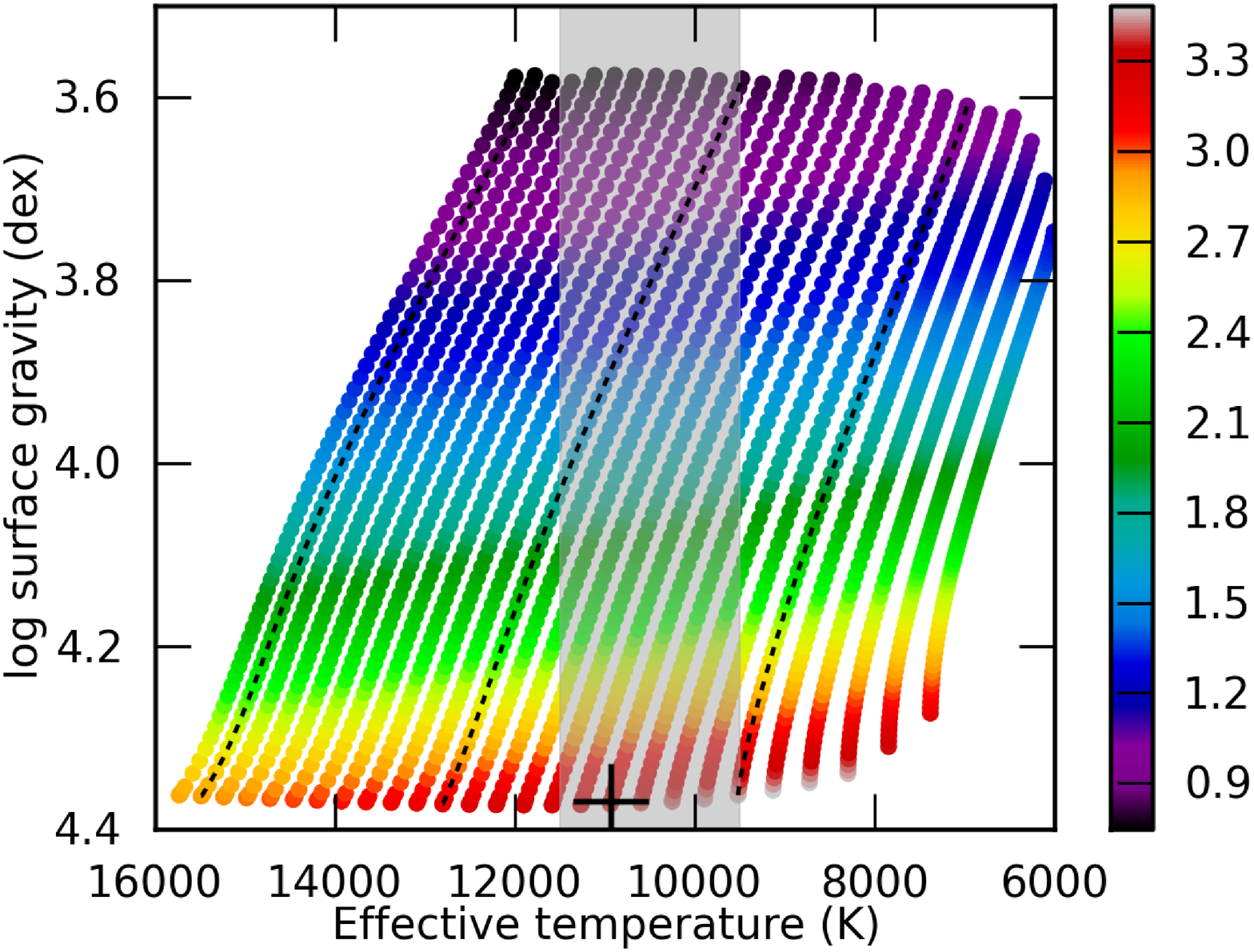} 
\includegraphics[width=0.98\columnwidth]{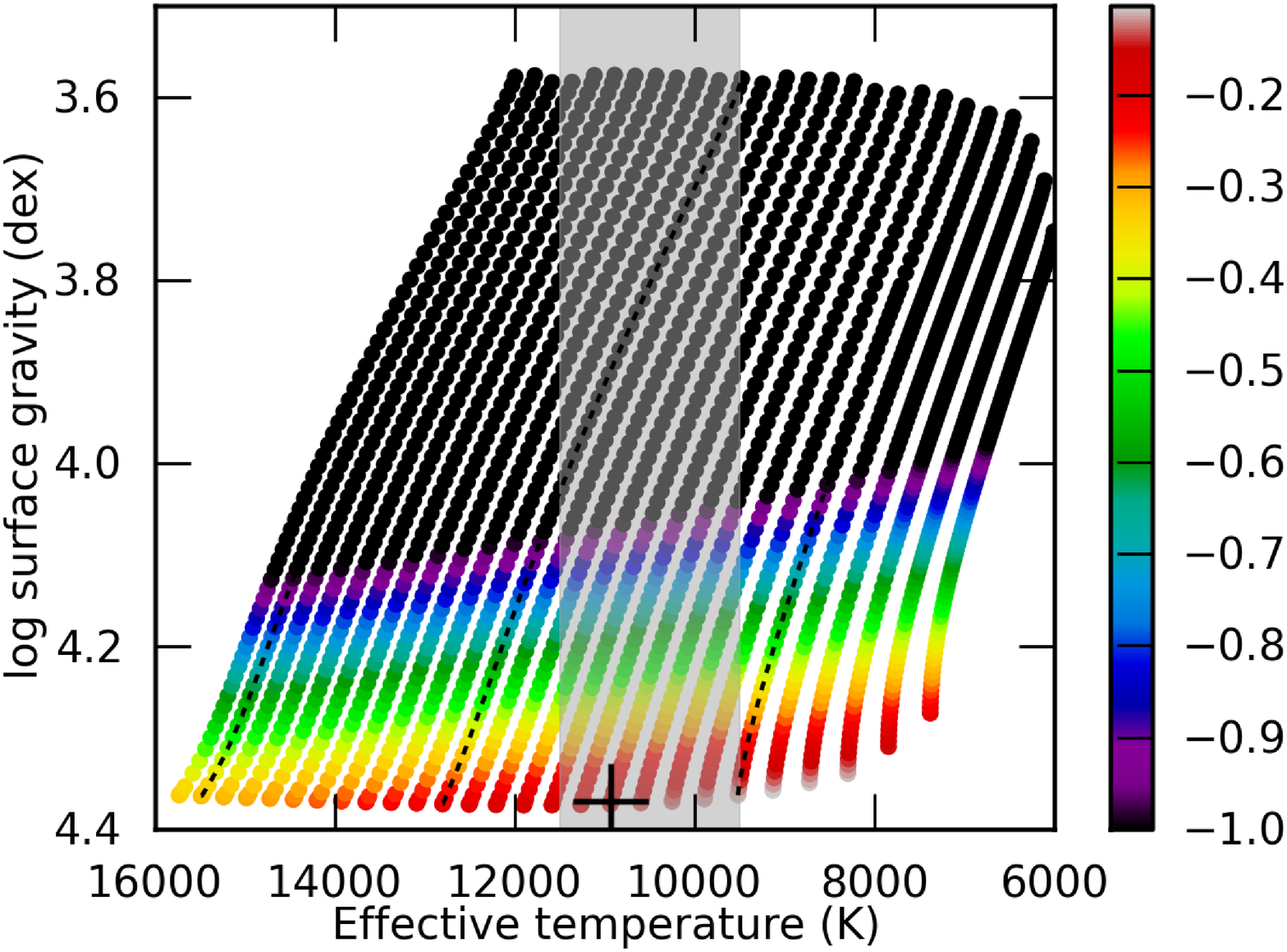} 
\caption{Critical rotation velocities and frequencies, and their relation to stellar parameters for models between 1.5 and 4.1\,M$_\odot$. The grey band shows the range of $T_\mathrm{eff}$ matching the observed spectra, the $+$ sign shows the location of the best matching model in km\,s$^{-1}$ (see text). \emph{Top left panel}: critical velocities on the main sequence for CL\'ES models with $\alpha_\mathrm{ov}=0.2$. The black line shows the line of models matching the observed $v_\mathrm{eq} \sin i$ with $i=90^\circ$. The grey lines show the models reaching their critical rotation frequency for different inclination angles for the observed $v_\mathrm{eq} \sin i$ ($i=40^\circ, 45^\circ, 60^\circ, 75^\circ$ and $90^\circ$). The colours represent the values of different of the following quantities. \emph{Top right:} $R_\mathrm{eq}$ for zero rotation ($R_\odot$). \emph{Bottom left:} Critical rotation frequency (d$^{-1}$). \emph{Bottom right:} Necessary differential rotation rate $\alpha$ for a model to reach $f_1$ at the pole with $\Omega_\mathrm{eq}=\Omega_\mathrm{crit}$.} 
\label{fig:coolbstars:HD174648:critical} 
\end{figure*}

The hypothesis that the photometric variability is due to rotational modulation implies a high rotational frequency equal to $f_1$ and $f_2$ at the spots' latitudes. In fact, a confrontation between the break-up frequencies for a grid of CL\'ES \citep{scuflaire2008} models \citep[from ][]{degroote2010b}, shows that $f_2$ exceeds the break-up frequency for all models near the location of HD\,174648 in the Kiel diagram (Fig.\,\ref{fig:coolbstars:HD174648:critical}). Nonetheless, if we account for the possibility of differential rotation, which is hinted at in the spectral profile of Mg\,\textsc{ii}\,4481\,\AA\ (Sect.\,\ref{sect:coolbstars:fundamental_parameters}), then the stellar surface at higher latitudes could possibly rotate faster without exceeding the local break-up velocity. In Fig.\,\ref{fig:coolbstars:HD174648:critical}, the minimum required amount of differential rotation $\alpha$ for a model to reach a rotation rate of $f_1$ at the pole with $\Omega_\mathrm{eq}=\Omega_\mathrm{crit}$ is shown. Here, $\alpha$ is defined via
\begin{equation}\Omega(\theta) = \Omega_\mathrm{eq} (1-\alpha\sin^2\theta).\label{eq:methodology:differential_rotation}\end{equation} 

In the case of polar acceleration $\alpha$ is negative and a minimal differential rotation rate is equivalent to a maximal $\alpha$, thus we search for 
\[\alpha_\mathrm{max} = \frac{\Omega_\mathrm{crit}-f_1}{\Omega_\mathrm{crit}}.\] 
By placing $f_1$ at the pole, we force the differential rotation law to be as close to uniform rotation ($\alpha=0$) as possible. We see that, unless the star is located close to the ZAMS, unrealistically high differential rotation rates have to be reached \citep[cf. rates up to $\sim$$-30\%$ reported by ][]{reiners2003}. On the other hand, \citet{reiners2004} have shown that the differential rotation rate increases with effective temperature. The hottest stars in their sample have an effective temperature 3\,000\,K lower than HD\,174648. Therefore, we restrict the different rotation rate to $\alpha>-40\%$ in the spot modelling in Sect.\,\ref{sect:solutions}.
 
In contrast to the rotation frequency, the projected equatorial rotational velocity is dependent on the inclination angle of the star. The only constraint on the spot model imposed by the observations is that we cannot have a model with too low a surface gravity, to avoid the stellar radius to become too large, and, given the observed rotation frequency, the measured $v_\mathrm{eq}\sin i$ to exceed the break-up velocity (Fig.\,\ref{fig:coolbstars:HD174648:critical}). In the following, we check if all data sources can be reconciled with the hypothesis of a differentially rotating spotted early-type star. After defining the general model, we will search for and discuss possible solutions. 
 
\subsubsection{The model} 

At a given time, the integrated intensity $S$ of the surface of a spotted star, with respect to its unspotted intensity $S_0$, can be approximated by \citep{lanza2003} 
\begin{equation}S = S_0 \mathcal{C} \sum_{n:\mu_n\geq0} A_n\mu_n h(\mu_n) \left[(c_s-1)+Q(c_f + c'_f\mu_n -1)\right] + S_0. \label{eq:methodology:spotmodel}\end{equation} 
The parameter $A_n$ represents the relative surface area of the $n$th spot, which is assumed to be circular, $\mu_n=\cos\psi_n$ is the cosine of the limb angle of the spot, and $\mathcal{C}$ is the unspotted specific intensity connected to the limb darkening function $h(\mu_n)$. The parameters $c_s$ and $c_f$ are the starspot and faculae contrast parameters, which determine how bright the spot is relative to the surroundings. For the faculae, an extra contrast coefficient $c_f'$ is introduced because the brightness of faculae is dependent on the viewing angle. The ratio between the area of the faculae and the starspots is fixed by $Q$, where $Q=0$ means no faculae. We chose to fix all spot parameters to the solar values even though the physical circumstances in the atmospheres are quite different for B  stars. The simplicity of our model justifies this choice, as a posteriori tests have shown, because the influence of higher contrast parameters, for example, can be mimicked by larger spots. In this respect, the physical reality of the formalism we used to model the spots is less important than the morphology it introduces in the light curve. Since our model treats spots as point sources, they cannot interfere with each other (i.e. combine or deform), and thus larger spots could also be interpreted as darker spots. For the Sun, the starspot contrast parameter takes the value $c_s=0.67$. The facular contrast coefficients are $c_f=1.115$ and $c_f'=-0.115$, and the ratio between the area of the faculae and starspot area is $Q=10$. 
 
In the adopted approach, the limb darkening law is approximated by the quadratic law 
\[h(\mu_n) = a_0 + a_1\mu_n + a_2\mu_n^2.\] 
The integration of Eq.\,\ref{eq:methodology:spotmodel} is done over the visible part of the star, defined as $\mu_n\geq0$, where 
\[\mu_n = \cos n \sin\theta_n + \sin i\cos\theta_n\cos(\phi_n+\Omega(\theta_n) t - L_0).\] 
The inclination of the star is denoted by $i$ (by convention, $i=0$ is pole-on, $i=\pi/2$ is equator-on), which is the opposite orientation of the latitude $\theta_n$ ($\theta_n=0$ represents a spot on the equator). $L_0$ is the epoch angle where $L_0=0$ means that a spot with coordinates $(\theta_n,\phi_n)=(0,0)$ is at the central meridian of the visible disk when $t=0$. Finally, the angular velocity $\Omega$ is in general dependent on the latitude via Eq.\,\ref{eq:methodology:differential_rotation}.
 
Spot modelling using only the light curve as a source of information is a degenerate problem. It is difficult to apply traditional minimization algorithms to the equations to find the correct parameters, because there are many local minima in the parameter space. Also, it is difficult to find good starting values to initialize the fit. In the case of a low number of spots, a genetic minimization algorithm can be a valuable alternative fitting method. These kinds of algorithms are generally better at exploring large parameter spaces and finding global minima, but require more evaluations. To facilitate the search we fix as many parameters as possible beforehand (e.g., the spot contrast ratios). Nevertheless, the outcome should be regarded as a \emph{possible} solution, not a \emph{unique} one. 
 
The genetic algorithm to fit Eq.\,\ref{eq:methodology:spotmodel} comprises the following steps: 
\begin{enumerate} 

 \item Initialize a sample consisting of thousands of sets of parameters (individuals), each representing a spot model. Choose the values of the parameters randomly within appropriate search ranges. 

 \item Evaluate all spot models, and compute the residuals with respect to the original light curve. For each solution calculate its fitness, i.e. a number representing the quality of the fit. If a fit is good, it should have a high fitness. We choose the fitness value to be the reciprocal of the $\chi^2$ statistic. 

 \item Rank the individuals in the sample according to their fitness, and select the $n$ best ones. This selection process picks a random fitness threshold between the worst and best individual, and retains only those individuals with a fitness higher than the threshold. 

 \item Keep the best individuals in the sample, and generate offspring to replace the bad individuals. Offspring are generated from two good individuals: sometimes switching some values, sometimes just copying them (random random-point crossover). Also the parents are randomly chosen, meaning that some of the good individuals may not generate any offspring, while others have many. Allow a random perturbation of all parameters (mutation), but make it small in the beginning, and slowly raise the perturbation during the evolution. 

\end{enumerate} 
Now reiterate steps (1)-(4) to breed successive generations, until a stop criterion is reached. This can be either after a certain number of generations or at a point where the sample fitness does not increase any more. 
 
\subsubsection{Solutions} \label{sect:solutions}

The free parameters in the fit describing the star are the equatorial period $P_\mathrm{eq}\in[0.26,0.43]$\,d and the differential rotation rate $b=(\Omega_\mathrm{p} - \Omega_\mathrm{eq})\in[0,1.5]$\,d$^{-1}$. Two infinitely lasting spots are assumed to be located on the same hemisphere, with longitude $\lambda_i\in[0,2\pi]$, latitude $\theta_i\in[0,\pi/2]$ and size $A_i$ between 0.001\% and 0.2\% of the surface area. In the genetic fit, the population consists of 15\,000 individuals, and by default, 100 evolution steps are made. On top of the random random-point crossover breeding, mutation is added to all breeding phases in the genetic fit. This is applied to all parameters $a_j$ in the form of normally distributed random noise with standard deviation $\sigma=\mu \alpha^k |a_j|$, where $\mu=1\times10^{-9},\alpha=1.2$ and $k$ is the generation number. This results in a power-law increase of the mutation parameter from virtually zero during the first evolution stages, to $\sim$10\% at the end. The last stages can then be used to check the influence of small perturbations on the fit parameters to the fit statistic. 
 
To investigate the uniqueness and convergence of the genetic fitting algorithm, we have performed the fit several times. From this exercise, we see that many combinations of parameters give a good fit. For example, with a low differential rotation rate, the frequency spacing $\Delta f$ can be recovered by placing the spots at high latitudinal separation. A higher differential rotation rate forces the separation to be smaller. This does not mean that we cannot formulate any constraints. Indeed, we have to cross-correlate the solutions from the fit with the grid of stellar models, the measured projected equatorial rotation velocity and the spectral profiles. Moreover, the differential rotation rate has to be realistic. 
 
To incorporate all collected restrictions on the star, we devised the following scheme: 
\begin{enumerate} 

\item For every model in the grid of evolutionary tracks we know the mass and radius in absence of rotation. Thus, we can compute the critical equatorial rotation frequency $\Omega_\mathrm{crit}$ (uniform rotation approximation) and compare it with the rotation frequency $\Omega_\mathrm{eq}=1/P_\mathrm{eq}$ retrieved from each separate spot model fitted with a realistic differential rotation rate ($\alpha>-40\%$). If $\Omega_\mathrm{crit}<\Omega_\mathrm{eq}$, we consider the fit to be incompatible with the stellar model under consideration. 

\item Next, we can derive the equatorial radius and check if it is compatible with the measured $v_\mathrm{eq}\sin i$ and the inclination angle from the spot model fit (i.e., we require that the difference cannot exceed 20\,km\,s$^{-1}$). We find that all valid models are located near the ZAMS. 

\item Finally, we compute synthetic spectra for a selection of $\sim$50 different valid models, varying in surface gravity, effective temperature, inclination angle and rotation frequency. Here, we take into account surface distortion using the rigidly rotating Roche approximation \citep{cranmer1995} ($\tilde{\Omega}=\frac{\Omega}{\Omega_\mathrm{crit}}$) 
\begin{equation}R(\theta) = \frac{3R_p}{\tilde{\Omega}\cos\theta} 
\cos\left(\frac{\pi + 
\arccos(\tilde{\Omega}\cos\theta)}{3}\right)\label{eq:methodology:distorted_surface},\end{equation} 
but we add a differentially rotating velocity field in compliance with the parameter resulting from the spot model fit. We note that this is not entirely consistent with the Roche approximation, but the additional surface distortion effects due to differential rotation are small. After discretising the surface of the star in $400\times200$ surface elements in longitude, and latitude, respectively, we compute the synthetic spectra in the following way: First, the vectors normal to the surface defined in Eq.\,\ref{eq:methodology:distorted_surface} are projected onto the line-of-sight. Then, a velocity field is constructed or derived and the magnitude of the velocity field in the line-of-sight is calculated. For each surface element, the local gravity $\log g(\theta)$ and temperature $T(\theta)$ are derived and the corresponding line profile is taken from the library of line profiles. The total line profile $J$ is then calculated by summing all shifted line profiles $I$ for each surface element, weighted with the amount of flux emitted in the line-of-sight $A$. The flux is calculated according to the \citet{vonzeipel1924} theorem, and corrected for the element area and limb darkening $h_\lambda$, which is derived from a library of specific intensities, using again the local gravity and temperature, 
\begin{equation}J = \sum_{\theta,\phi}\left[\delta(\theta,\phi,z)A(\theta,\phi,z)h_\lambda(\mu,T_l,\log g_l)I(v,T_l,\log g_l)\right].\label{eq:methodology:synthetic_lineprofiles}\end{equation} 
\end{enumerate} 
 
\begin{figure*}                   
\centering\includegraphics[width=2\columnwidth]{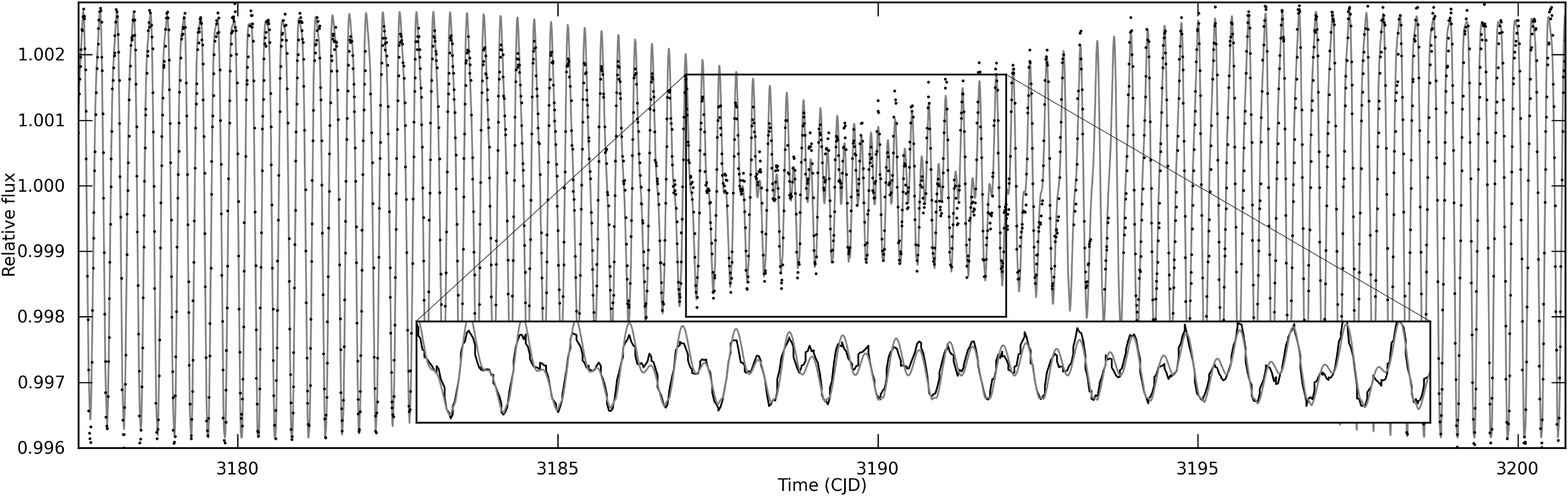} 
\includegraphics[width=2\columnwidth]{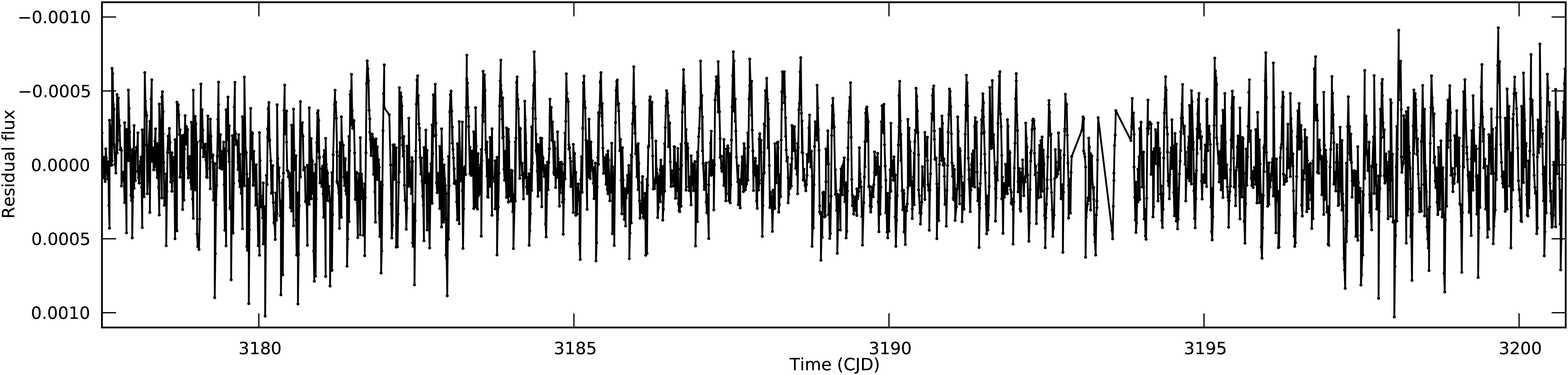} 
\caption{Spot model fit and residuals. \emph{Top panel:} Black dots and lines show the observed CoRoT light curve, corrected for discontinuities. In grey, the best fitting spot model with parameters listed in Table\,\ref{tbl:coolbstars:HD174648:spotmodel} is shown. The inset is a zoom on the middle part of the light curve. \emph{Bottom panel:} Residuals after subtracting the spot model.} 
\label{fig:coolbstars:HD174648:spotmodel} 
\end{figure*}

\begin{figure} 
\centering \includegraphics[width=\columnwidth]{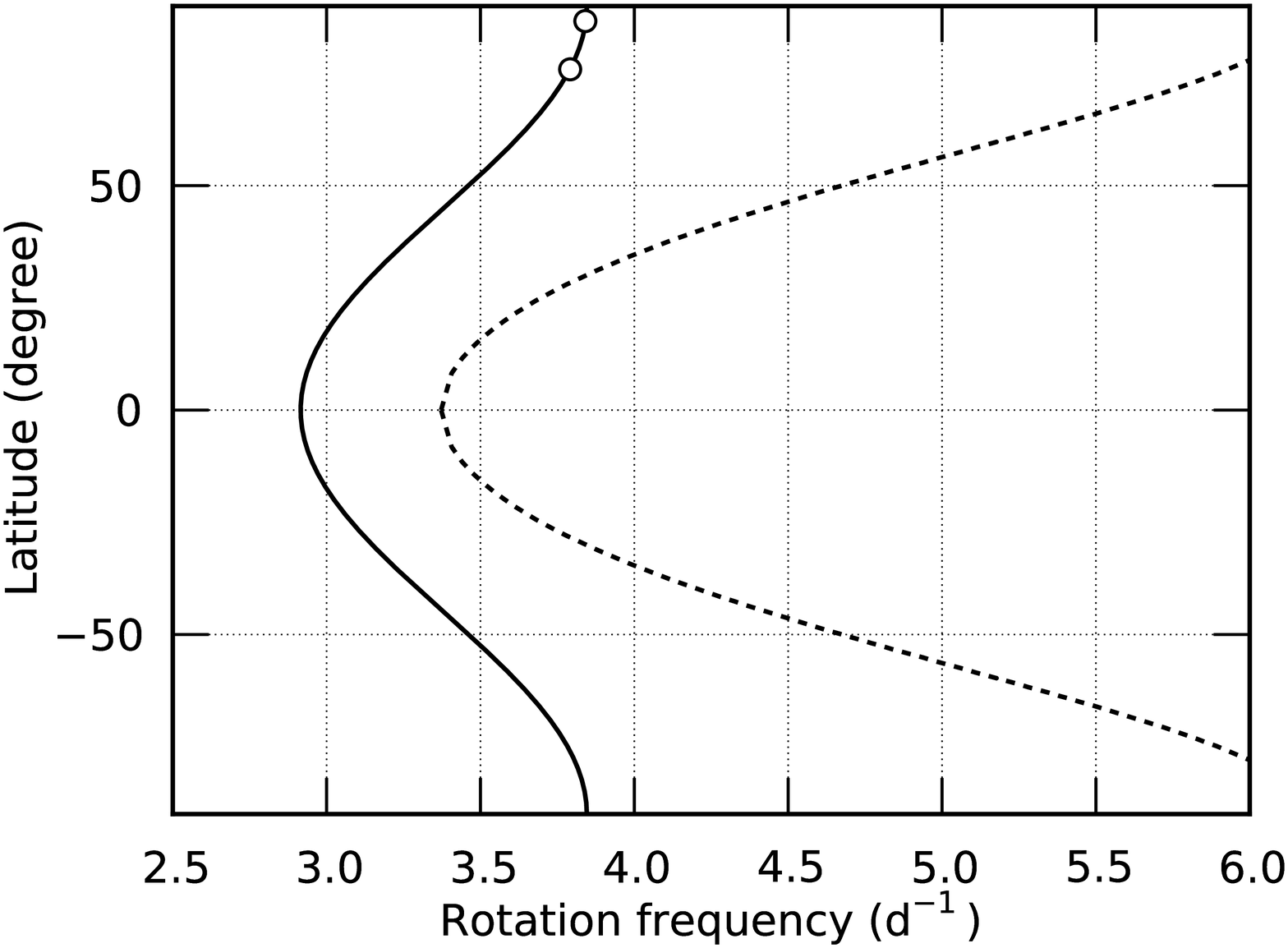} 
\caption{Differential rotation law for the best fitting spot model (solid line). In dashed lines, the breakup rotation frequency is shown as a function of latitude. The locations of the two spots are shown as white circles.} 
\label{fig:coolbstars:HD174648:rotlaw} 
\end{figure}

\begin{figure} 
\includegraphics[width=0.49\columnwidth]{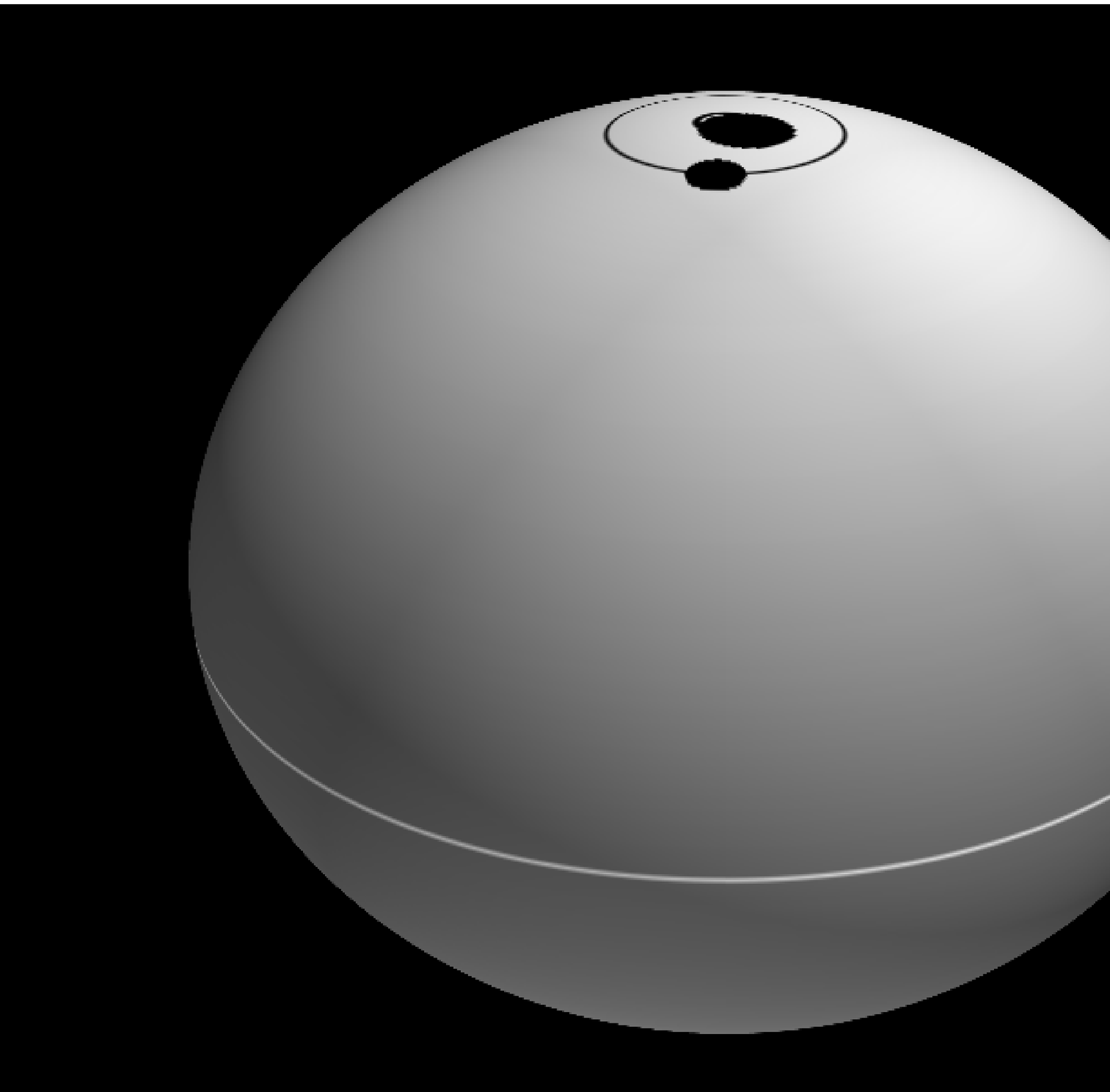} 
\includegraphics[width=0.49\columnwidth]{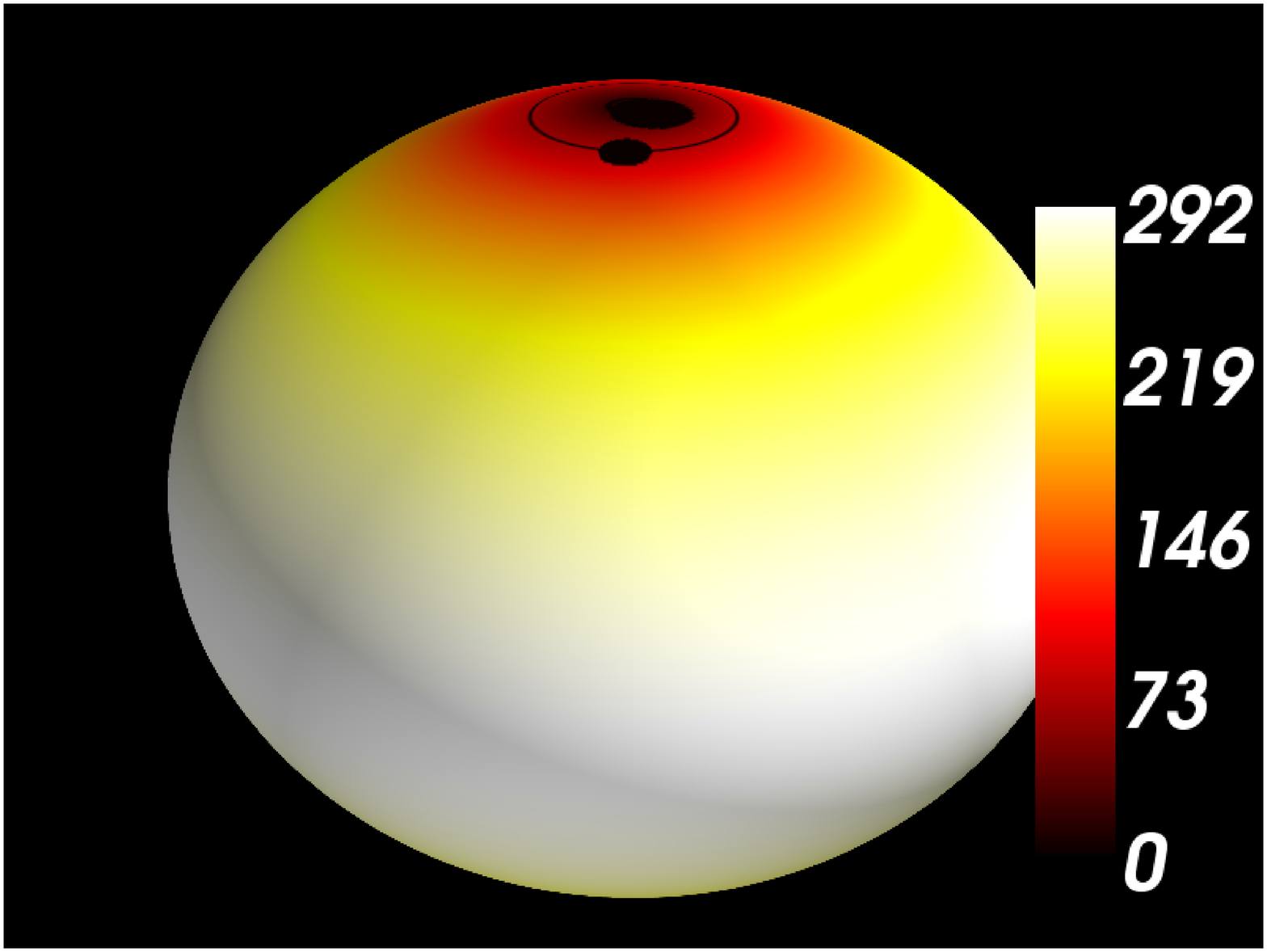} 
\includegraphics[width=0.49\columnwidth]{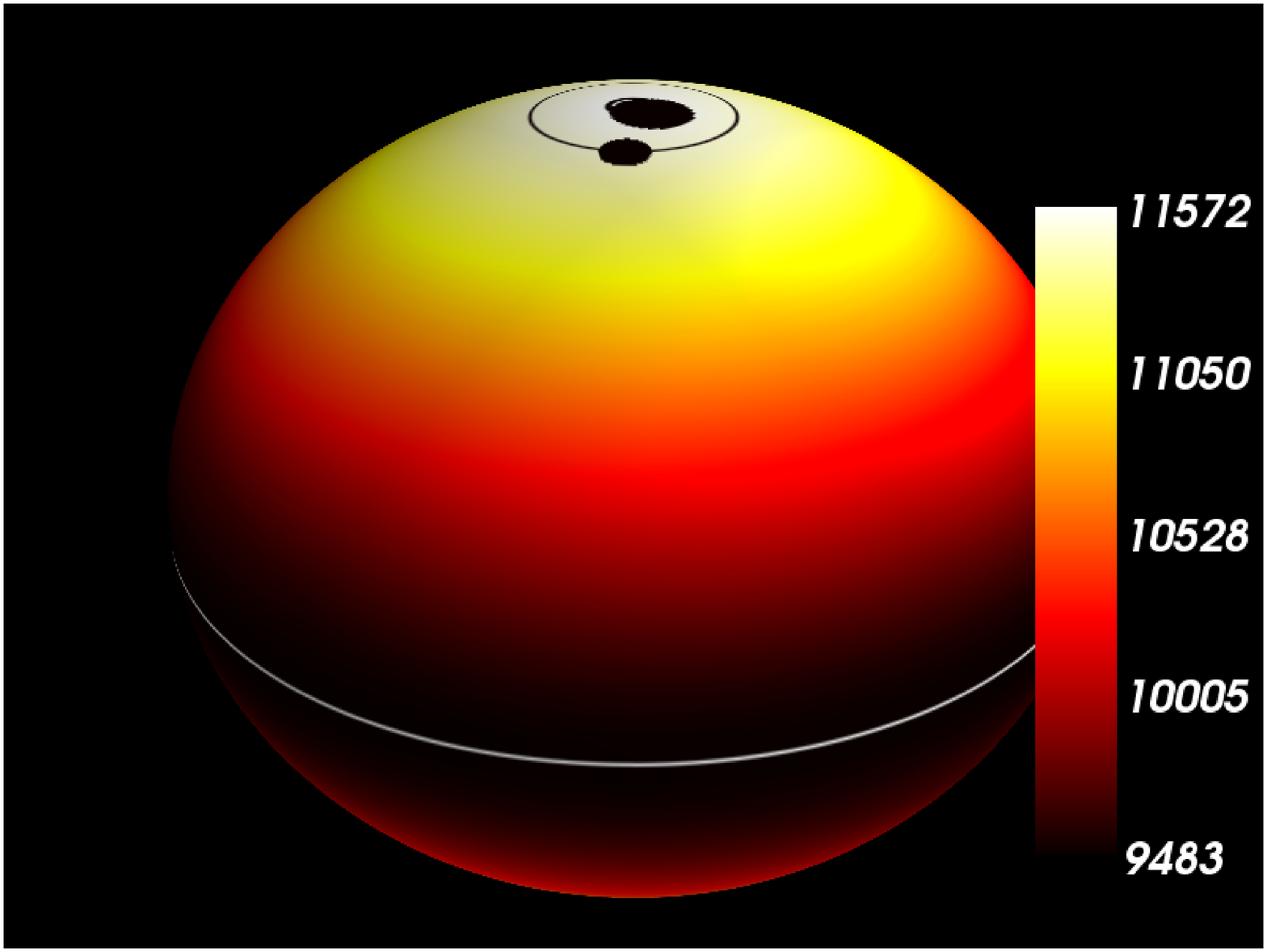} 
\includegraphics[width=0.49\columnwidth]{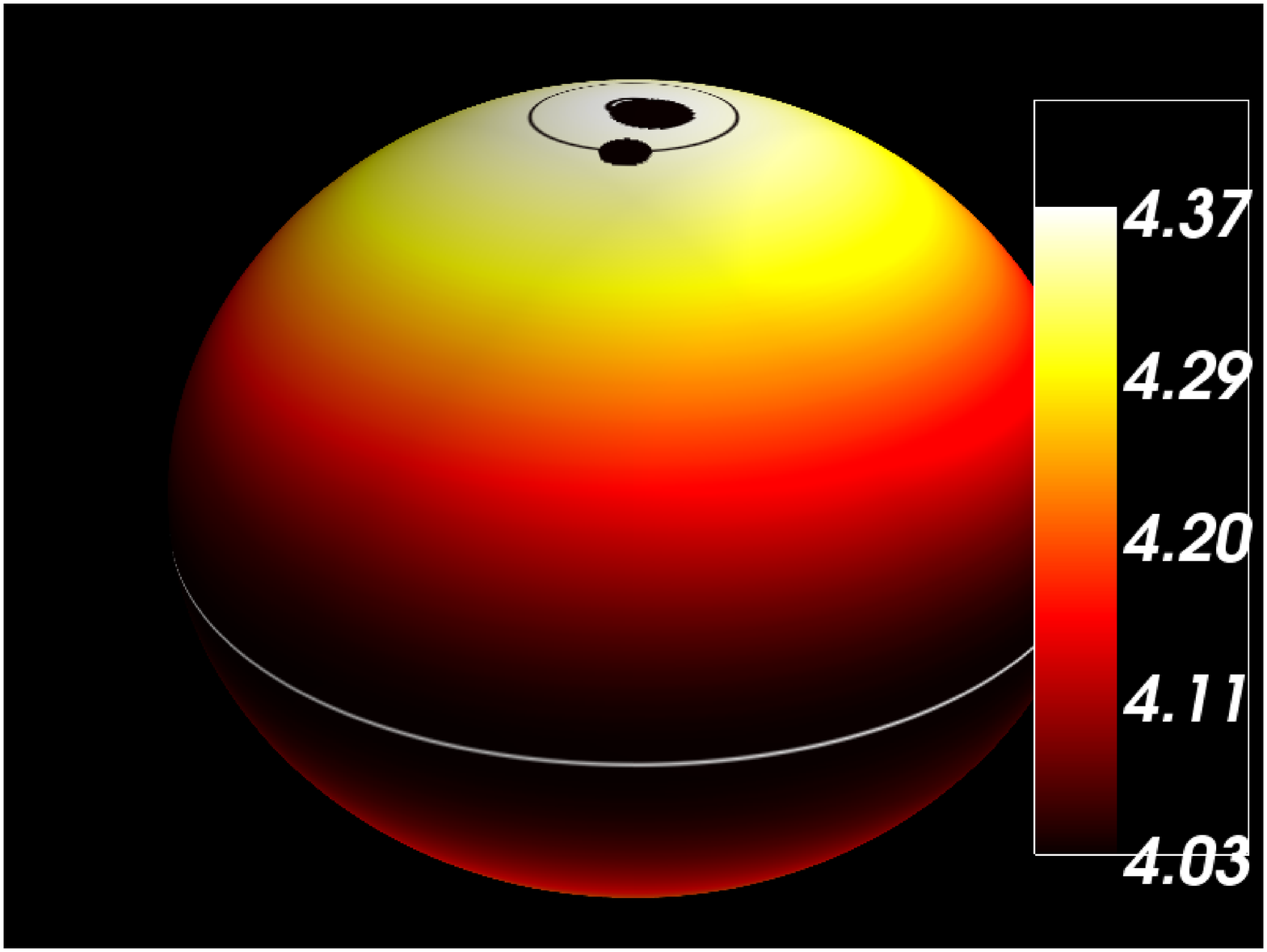} 
\caption{Stellar parameters across the rotationally distorted surface of the star, and location of the spots. The paths of the spots are shown in black lines. The equator of the star is shown as a white line. \emph{Top left:} emitted flux (linear scale). \emph{Top right}: rotation velocity (km\,s$^{-1}$). \emph{Bottom left:} effective temperature (K). \emph{Bottom right:} logarithmic surface gravity (cgs) (coloured version only online).} 
\label{fig:coolbstars:hd174648:3dmodels}   
\end{figure}

To compare the synthetic spectra with the observed spectra, we applied the same fitting algorithm as described in Sect.\,\ref{sect:coolbstars:fundamental_parameters} (Fig.\,\ref{fig:coolbstars:hd174648:fundpars}, red line). These line profiles are calculated by numerically summing the contributions of each surface element. For each surface element, the spectral features and intensity are derived from the local effective temperature and local surface gravity, thus taking gravity darkening effects into account. We note that the line profiles can differ more from one another than what would be expected from the difference in polar effective temperature alone. The rotation can lower the effective temperature at the equator by several thousand Kelvin, which can have a major effect on the shape of the line, even if the polar temperature changes only slightly. The parameters of the best matching model are listed in Table\,\ref{tbl:coolbstars:HD174648:spotmodel}, and the resulting light curve model is compared to the observations in Fig.\,\ref{fig:coolbstars:HD174648:spotmodel}. The rotation law is illustrated in Fig.\,\ref{fig:coolbstars:HD174648:rotlaw}, and the distribution of the flux, temperature, gravity and velocity on the surface are shown in Fig.\,\ref{fig:coolbstars:hd174648:3dmodels}. 
 
We compare the fit quality of the spotted model with the pulsation model via a reduced $\chi^2$ statistic, 
\[\chi_{\mathrm{red}}^2 = \frac{1}{N}\sum_\lambda\frac{(F_\mathrm{mod,\lambda} - F_\mathrm{obs,\lambda})^2}{\sigma^2}.\] 
Here, $\sigma$ is determined as $\sqrt{2}$ times the standard deviation of the differenced residual light curve after subtracting the pulsation model resulting from a full frequency analysis. Considering only two frequencies (7 free parameters), we find $\chi_1^2=9.14$. Adding one harmonic of the main frequencies, we end up with $\chi_2^2=1.26$ (11 free parameters). The spot model results in $\chi_3^2=2.61$ (10 free parameters). We conclude that, purely on the basis of fit quality, neither the spot hypothesis nor the pulsation hypothesis can be excluded. 
  
\begin{table} 
\centering 
\caption{Spot model parameters for HD\,174648.} 
\begin{tabular}{lc} 
\hline 
\hline\\ 
Parameter & Value\\\hline 
 $M$ ($M_\odot$) & $2.4\pm0.1$ \\ 
 $R_\mathrm{pole}$ ($R_\odot$) & 1.7\\
 $T_\mathrm{eff,pole}$ (K) &  $10\,950\pm500$  \\ 
 $\log g_\mathrm{pole}$ (dex)       & $4.37\pm0.2$      \\ 
 $P$ (d)              & $0.34 \pm 0.02$   \\ 
 $\alpha $            & $-0.32\pm 0.03$\\ 
 $\Omega_\mathrm{eq}$ ($\Omega_\mathrm{crit}$)  & 0.87\\
 $\Omega_\mathrm{pole}$ ($\Omega_\mathrm{crit}$)  & 1.14 \\ 
 $v_\mathrm{eq}\sin i$ (km\,s$^{-1}$) & 283\\ 
 $i$ ($^\circ$)       & $76.3\pm4$      \\ 
 $\theta_1$ ($^\circ$)    & $75.9\pm4$\\ 
 $\theta_2$ ($^\circ$)    & $86.6\pm4$\\ 
 $s_1$ (surface fraction) & $0.0135\pm0.005$\\ 
 $s_2$ (surface fraction) & $0.0385\pm0.003$\\ 
 $s_1'$ (surface fraction) & 0.0010 ($r\approx4^\circ)$\\ 
 $s_2'$ (surface fraction) & 0.0049 ($r\approx3^\circ)$\\ 
\hline\hline 
  
\end{tabular} 
 
\label{tbl:coolbstars:HD174648:spotmodel} 
\end{table} 
   
Next, we checked if the location of the spots and the star's inclination angle resulting from the best fitting model are compatible with the amplitude of the variability. Because of CoRoT's single-colour white light photometry, we cannot determine the spot temperature, and because of the disc integration, we cannot infer the morphology of the spots. For simplicity we therefore assumed the spots to be perfectly circular and totally black (i.e., blocking all radiation and not experiencing any faculae). We then placed these dark patches on the distorted surface of the model star, and computed the minimum size of the spots to induce $\sim$0.18\% and $\sim$0.13\% flux variations. The largest spot at 86.6$^\circ$ is compatible with a spot radius of $\sim$$4^\circ$ (parameter $s_1'$ in Table\,\ref{tbl:coolbstars:HD174648:spotmodel}), or a surface coverage of $\sim$0.1\%. The smallest spot at 75.9$^\circ$ should have a radius of $\sim$$3^\circ$ corresponding to a surface coverage of $\sim$$0.05\%$ (parameter $s_2'$ in Table\,\ref{tbl:coolbstars:HD174648:spotmodel}). These are plausible values; the spots can be small compared to the surface area, and do not need to overlap. In fact, a reduction of 10\% of the original flux at the spot locations can already induce the same variability with only a marginal increase of the spot size. 
These values are smaller than those resulting from the spot model fit (Table\,\ref{tbl:coolbstars:HD174648:spotmodel}, parameters $s_1$ and $s_2$), where we assumed solar values for all spot parameters. 
 
\subsubsection{Discussion} 
 
The two dominant frequencies in the CoRoT light curve of HD\,174648 both have a clear second harmonic, which is a typical signature of spot-like features. The fact that we do not see a constant part in the phase shapes of the light curve folded on the dominant frequencies implies that the spots never fully disappear behind the stellar disk. This means that we either see the star nearly equator-on with two circumpolar spots, or that the star has a significant inclination with respect to our line of sight and the spots can be anywhere on the visible hemisphere. The first possibility is favoured by the high projected rotational velocity derived from the Mg\,\textsc{ii}\,4481\AA\ line profile. The hypothesis of circumpolar spots is also supported by the rotation frequency at the spot latitudes. The frequencies $f_1$ and $f_2$ are too high to be equatorial rotation frequencies, since they are higher than the break-up frequency for any realistic stellar model compatible with the fundamental parameters of the star. Fast rotating circumpolar spots are then compatible with an equatorial rotation frequency below break-up velocity and a high differential rotation rate with polar acceleration. A sign of high differential rotation with polar acceleration is also found in the shape of the Mg\,\textsc{ii}\,4481\AA\ line profile. According to \citet{reiners2002b}, polar spots can mimic differential rotation with polar acceleration in spectroscopic line profiles, but this degeneracy is lifted based on the light curve morphology. Uniform rotation would remove the beating pattern since the two spots would have the same rotation period.
 
We have conceptually proven the possibility of the two-spot model to explain the light curve of HD\,174648, and we have shown that the resulting parameters are compatible with models of stars close to the ZAMS. The solution is in agreement with the observed spectral lines. Four key features have to be explained to make the spot hypothesis plausible from a purely physical vantage point: rapid rotation, a high differential rotation rate, polar acceleration and the existence of circumpolar spots. Rapid rotation is most prominent in stars of spectral type A and B, of which the former have the highest fractional rotation velocities on average, and the latter the highest absolute rotation velocities. The polar acceleration is compatible with rapid rotation in stars without a significant convective outer layer \citep{clement1969}: convection with viscous braking implies equatorial acceleration (e.g., the Sun), but in the absence of significant convective layers, radiative braking takes over as the most important braking mechanism, and results in polar acceleration. Due to the high effective temperature of HD\,174648, we expect this star only to have a shallow convective layer, if any, and thus rapid rotation and polar acceleration are indeed possible. \citet{zorec2011} show that rapid rotation can lead to differential rotation in early type stars, and \citet{hussain2002} and \citet{isik2007} show that rapid rotation can lead to a concentration of spots at polar latitudes, because of an efficient transport of flux from lower latitudes via meridional circulation. The same mechanism can also transport the magnetic field to higher latitudes and induce strong polar magnetic fields, benefiting large, long-lived spots. Finally, \citet{hussain2002} shows that the differential rotation rate increases with stellar mass, though we have to be careful not to overinterpret her simulations, since they were carried out for cooler stars (masses up to $1.7$\,M$_\odot$). On the other hand, \citet{reiners2004} have analysed a sample of A stars, and found evidence for a high differential rotation rate of $\sim$$30\%$ in the earliest-type object in their sample \citep[HD\,60555, A6V, $v\sin i=114$\,km\,s$^{-1}$; ][]{royer2007}. 
 
Nonetheless, some problems remain. If only a very small convective region exists, how can a magnetic field be generated strong enough to create spots \citep[cf. ][]{cantiello2010}? Even though the above literature studies are supportive of our interpretation, a strong correlation between the presence of magnetic fields and \emph{low} rotation rate was previously observed \citep{briquet2007}, which is not in agreement with the fast rotation of HD\,174648. Magnetic fields and/or spots have only been detected in the slowly rotating chemically peculiar Ap/Bp stars, HgMn stars \citep{wolff1983,briquet2001,briquet2007,alecian2009,briquet2010b}, and in SPBs \citep{hubrig2009}. Thus, if a magnetic field is present, it should not only be weak, but also fast-evolving and probably non-dipolar, to allow for the existence of close circumpolar spots and differential (fast) rotation. This scenario is not reconcilable with what has been found for the magnetic Ap/Bp stars \citep{shorlin2002,auriere2007}. Yet, rotational modulation has been proposed for line profile variations in other rapidly rotating stars, sometimes in combination with pulsations \citep[e.g., the $\beta$\,Cep binary $\kappa$\,Sco, ][]{uytterhoeven2005}. Another explanation for the spots is to invoke inhomogeneities in surface chemical abundances. In this case, the problems linked to Bp-like magnetic fields vanish, since these inhomogeneities are observed in stars for which no large-scale or strong magnetic fields have been detected \citep[e.g., ][]{wade2006, auriere2010}.

A final remark concerns the possibility of contamination of the CoRoT light curve by variability of a cooler star in the same field of view. A foreground spotted K5V star can induce an amplitude of 1800\,ppm in the light curve. At the distance of HD\,174648 or further away, the spot amplitude has to be larger than 0.2\,mag to cause the observed variability, which is unlikely. The detected frequency, however, would imply an equatorial rotation velocity of 170\,km\,s$^{-1}$, which can only be obtained in rare cases of a tidally locked compact binary (e.g., a K5V star around a white dwarf with a semi-major axis of $\sim$$2R_\odot$).

\section{The pulsating stars} 
\begin{figure*} 
\includegraphics[width=2\columnwidth]{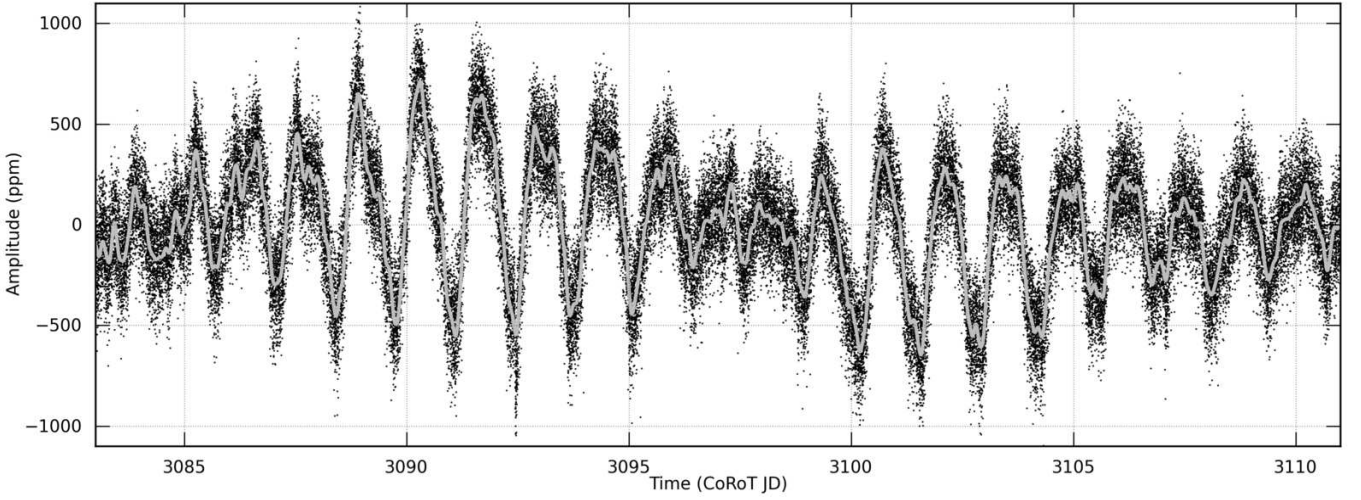} 
 \caption{Excerpt of the CoRoT light curve of HD\,170935, showing similar variability as HD\,174648 (grey line shows a fit to the light curve, black dots represent to CoRoT data).}
\label{fig:hd170935} 
\end{figure*}

In terms of light curve morphology, features in the frequency spectrum (e.g., the presence of harmonics) and rotation rate ($v_\mathrm{eq}\sin i=271\pm16$\,km\,s$^{-1}$), HD\,170935 resembles HD\,174648 (Fig.\,\ref{fig:hd170935}). The surface gravity suggest that it is slightly more evolved, frequency values themselves are considerably lower ($\sim$$0.70$\,d$^{-1}$ compared to $\sim$$3.8$\,d$^{-1}$), and the number of frequencies higher with respect to HD\,174648. Explaining this light curve in terms of spots would require a complex pattern of numerous spots. Nonlinear gravity mode pulsations are therefore a more likely cause of variability, making HD\,170935 a fast rotating and cool SPB star, just as HD\,121190 \citep{aerts2005}.

The four other cool B stars in the CoRoT sample (Figs\,\ref{fig:hd181440}-\ref{fig:hd46179}) contain fewer frequencies than the two stars discussed previously. Moreover, at low amplitudes, residual instrumental effects can be important, diminishing the number of detected frequencies even further. Except for the frequencies with very high S/N, we assume that the bulk of the low frequencies are related to instrumental noise. The only star in the remainder of the sample with a clear multiperiodic frequency spectrum is HD\,181440 (B9III). The high S/N frequencies and candidate combination frequencies are listed in Table\,\ref{tbl:coolbstars:hd181440}. The presence of a multiplet around $f_1=0.44414\pm0.00014$\,d$^{-1}$ (containing $f_4,f_5$ and $f_6$) hints for existence of variable amplitudes, frequencies, or phases, but a beating pattern cannot be ruled out. No significant harmonics are extracted. The possibility of a multiplet is excluded from the low value of the spacing and the projected rotational velocity. The frequencies $f_2$ and $f_3$ are isolated and therefore assumed to be pulsations.

\begin{table*}[htb] 
\caption{ Frequency analysis of the CoRoT light curve of HD\,181440, after correcting for discontinuities. Only the frequencies with a $S/N>20$ are shown, and candidate combination frequencies.} 
\centering \begin{tabular}{lrrrrl} 
\hline\hline 
 \multicolumn{1}{c}{ID}
& \multicolumn{1}{c}{$f$}
& \multicolumn{1}{c}{$a$}
& \multicolumn{1}{c}{$\phi$}
&  \multicolumn{1}{c}{S/N} &\\ 
        
& \multicolumn{1}{c}{ d$^{-1}$}
& \multicolumn{1}{c}{ppm}
& \multicolumn{1}{c}{rad/2$\pi$}
&      &\\\hline 
$f_1$ & $ 0.444136 \pm 0.000014 $ & $ 72.3 \pm 0.3 $  & $  0.432 \pm 0.004 $ & 39.58 & \\ 
$f_2$ & $ 0.620652 \pm 0.000013 $ & $ 69.6 \pm 0.3 $  & $ -0.332 \pm 0.004 $ & 40.50 & \\ 
$f_3$ & $ 1.166012 \pm 0.000017 $ & $ 47.0 \pm 0.2 $  & $  0.441 \pm 0.005 $ & 30.60 & \\ 
$f_4$ & $ 0.449199 \pm 0.000024 $ & $ 33.7 \pm 0.2 $  & $ -0.354 \pm 0.007 $ & 22.91 & \\ 
$f_5$ & $ 0.430491 \pm 0.000026 $ & $ 27.9 \pm 0.2 $  & $  0.445 \pm 0.008 $ & 21.45 & \\ 
$f_6$ & $ 0.410365 \pm 0.000026 $ & $ 26.9 \pm 0.2 $  & $  0.053 \pm 0.007 $ & 22.11 & \\ 
$f_7$ & $ 0.543475 \pm 0.000034 $ & $ 20.2 \pm 0.2 $  & $ -0.241 \pm 0.010 $ & 16.89 & $f_3-f_2+0.001890$\\ 
$f_8$ & $ 0.180443 \pm 0.000055 $ & $ 12.2 \pm 0.2 $  & $ -0.189 \pm 0.016 $ & 11.66 & $f_2-f_1+0.003927$\\ 
$f_9$ & $ 0.174743 \pm 0.000097 $ & $  7.6 \pm 0.2 $  & $  0.219 \pm 0.028 $ &  7.20 & $f_2-f_4+0.003291$\\\hline 
\end{tabular} 
\label{tbl:coolbstars:hd181440} 
\end{table*}

In the frequency spectra of the two slowly rotating stars HD\,182198 (B9V) and HD\,49677 (B9), only one frequency is clearly present. The frequencies $f=0.13976\pm0.00001$\,d$^{-1}$ (in HD\,182198) and $f=1.3364\pm0.0009$\,d$^{-1}$ (HD\,49677), are good candidates for an SPB gravity mode and could therefore be probes for the borders of the SPB instability strip. The occurrence of monoperiodic SPBs, or of such pulsators with a highly dominant mode, is common \citep{decat2005}. Finally, no clear stellar frequencies were recovered from the moderate rotator HD\,46179. 
 
\begin{figure} 
\centering \includegraphics[width=\columnwidth]{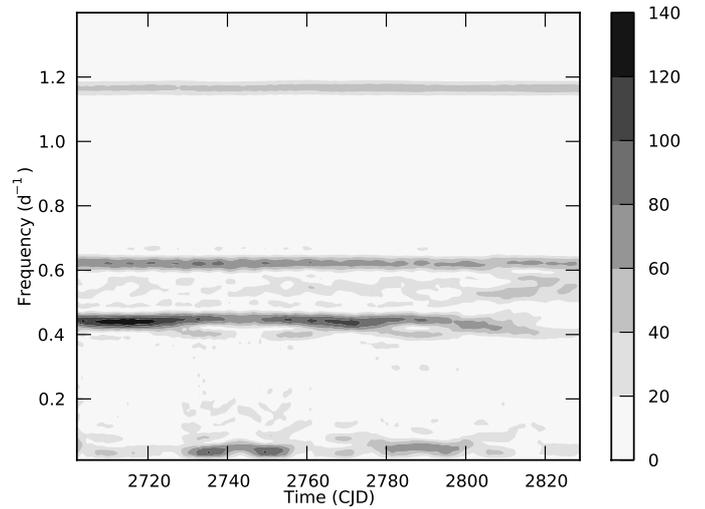} 
\caption{Short time Fourier transform of the CoRoT light curve of HD\,181440 (window length of 30\,d). The two highest frequencies are constant throughout the time series. The low frequency could result from rotational modulation, but a beating pattern cannot be excluded (amplitudes in ppm).}\label{fig:coolbstars:otherstars:stft} 
\end{figure} 

\begin{figure*} 
\includegraphics[width=2\columnwidth]{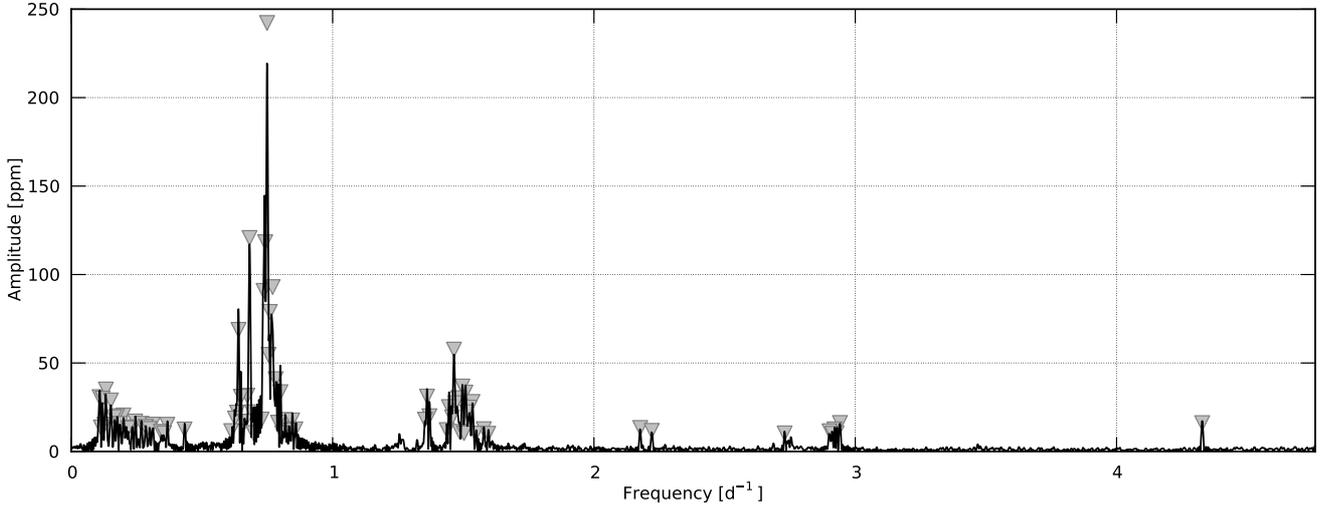} 
 \caption{Frequency spectrum of the CoRoT light curve of HD\,170935 above 0.1\,d$^{-1}$ (black). The significant frequencies and fitted amplitudes above 10 ppm are indicated in grey.}
\label{fig:per:hd170935} 
\end{figure*}

\section{Conclusions} 
 
The cool side of the theoretical SPB instability strip harbours many different types of stars, some of them exhibiting clear variability. SPB stars co-exist in this part of the HR diagram with Be stars, Bn stars, Bp stars and HgMn stars. The reasons for the differences among these classes of stars are not well understood, but rotation and magnetic fields must play important roles. We have argued that pulsations need not be the (only) source of variability for all these stars. Instead, rotational modulation is an alternative, at least for the fast rotators. For slow rotators or objects with low surface gravities, the rotation periods are of the same order as the expected pulsation periods, thus confusion occurs if only white-light photometric time series are available. Although time-frequency analyses hold the potential to distinguish between the two phenomena, the frequency spectrum of gravity modes is expected to be dense, thus long term continuous observations are needed to disentangle beating patterns from time-dependent behaviour. 
 
Despite the precision and duty cycle of single-band space-based photometry, we have shown that a unique and consistent description of late B star variability is difficult to obtain solely based on this source of data, even when multicolour photometry or spectroscopy are available to constrain the fundamental parameters such as the effective temperature and surface gravity of the star. We have also seen that a good determination of the rotation frequency is important. Additionally, multicolour photometric time series or spectroscopic time series are necessary to determine the origin of variability of these cool B stars, as spots and pulsations leave distinct signatures in the amplitude ratios for different photometric bandpasses, and in line profiles. 

We found evidence for the existence of differential rotation at the surface of one fast rotator of late spectral type B. This holds information on the internal rotation profile if we can pinpoint the cause of the nonrigidity, and can thus be a tool to probe the interior of massive stars. Only cool B stars were studied, but there is no reason to assume similar stars cannot exist with hotter temperatures. In fact, we could hope for the existence of SPB stars with inhomogeneous surface structures, which would give independent constraints on the rotational behaviour of these stars to compare with those obtained from asteroseismology. 
 
The origin of the spots remains an unanswered question. A strong, Bp-like magnetic field is incompatible with fast rotation and the need for a non-dipolar field to explain the close circumpolar spots to be of magnetic origin \citep{auriere2007}. On the other hand, a weak magnetic field and the existence of differential rotation could more easily explain the fast rotation, but would invoke the need for fast evolving magnetic fields to generate the spots. Chemical spots are a valuable alternative, since they appear not to need the presence of magnetic fields, though there is no consensus in the literature on an alternative underlying mechanism to generate them \citep{hubrig2001,wade2006,auriere2010}.

 
\begin{acknowledgements} 
The research leading to these results has received funding from the European Research Council under the European Community's Seventh Framework Programme (FP7/2007--2013)/ERC grant agreement n$^\circ$227224 (PROSPERITY), as well as from the Research Council of K.U.\,Leuven grant agreement GOA/2008/04 and from the Belgian PRODEX Office under contract C90309: CoRoT Data Exploitation. Based on observations obtained with the HERMES spectrograph, which is supported by the Fund for Scientific Research of Flanders (FWO), Belgium, the Research Council of K.U.\,Leuven, Belgium, the Fonds National Recherches Scientific (FNRS), Belgium, the Royal Observatory of Belgium, the Observatoire de Gen\`eve, Switzerland and the Th\"uringer Landessternwarte Tautenburg, Germany. 
\end{acknowledgements} 

\bibliographystyle{aa} 

\begin{thebibliography}{70}
\expandafter\ifx\csname natexlab\endcsname\relax\def\natexlab#1{#1}\fi

\bibitem[{{Aerts} {et~al.}(2006){Aerts}, {De Cat}, {Kuschnig}, {Matthews},
  {Guenther}, {Moffat}, {Rucinski}, {Sasselov}, {Walker}, \&
  {Weiss}}]{aerts2006}
{Aerts}, C., {De Cat}, P., {Kuschnig}, R., {et~al.} 2006, \apjl, 642, L165

\bibitem[{{Aerts} \& {Kolenberg}(2005)}]{aerts2005}
{Aerts}, C. \& {Kolenberg}, K. 2005, \aap, 431, 615

\bibitem[{{Aerts} {et~al.}(2004){Aerts}, {Lamers}, \&
  {Molenberghs}}]{aerts2004b}
{Aerts}, C., {Lamers}, H.~J.~G.~L.~M., \& {Molenberghs}, G. 2004, \aap, 418,
  639

\bibitem[{{Alecian} {et~al.}(2009){Alecian}, {Gebran}, {Auvergne}, {Richard},
  {Samadi}, {Weiss}, \& {Baglin}}]{alecian2009}
{Alecian}, G., {Gebran}, M., {Auvergne}, M., {et~al.} 2009, \aap, 506, 69

\bibitem[{{Asplund} {et~al.}(2005){Asplund}, {Grevesse}, {Sauval}, {Allende
  Prieto}, \& {Blomme}}]{asplund2005}
{Asplund}, M., {Grevesse}, N., {Sauval}, A.~J., {Allende Prieto}, C., \&
  {Blomme}, R. 2005, \aap, 431, 693

\bibitem[{{Auri{\`e}re} {et~al.}(2010){Auri{\`e}re}, {Wade}, {Ligni{\`e}res},
  {Hui-Bon-Hoa}, {Landstreet}, {Iliev}, {Donati}, {Petit}, {Roudier}, \&
  {Th{\'e}ado}}]{auriere2010}
{Auri{\`e}re}, M., {Wade}, G.~A., {Ligni{\`e}res}, F., {et~al.} 2010, \aap,
  523, A40+

\bibitem[{{Auri{\`e}re} {et~al.}(2007){Auri{\`e}re}, {Wade}, {Silvester},
  {Ligni{\`e}res}, {Bagnulo}, {Bale}, {Dintrans}, {Donati}, {Folsom},
  {Gruberbauer}, {Hui Bon Hoa}, {Jeffers}, {Johnson}, {Landstreet},
  {L{\`e}bre}, {Lueftinger}, {Marsden}, {Mouillet}, {Naseri}, {Paletou},
  {Petit}, {Power}, {Rincon}, {Strasser}, \& {Toqu{\'e}}}]{auriere2007}
{Auri{\`e}re}, M., {Wade}, G.~A., {Silvester}, J., {et~al.} 2007, \aap, 475,
  1053

\bibitem[{{Baade}(1989{\natexlab{a}})}]{baade1989a}
{Baade}, D. 1989{\natexlab{a}}, \aaps, 79, 423

\bibitem[{{Baade}(1989{\natexlab{b}})}]{baade1989b}
{Baade}, D. 1989{\natexlab{b}}, \aap, 222, 200

\bibitem[{{Baglin} {et~al.}(2006){Baglin}, {Michel}, {Auvergne}, \& {The COROT
  Team}}]{baglin2006}
{Baglin}, A., {Michel}, E., {Auvergne}, M., \& {The COROT Team}. 2006, in
  Proceedings of SOHO 18/GONG 2006/HELAS I, Beyond the spherical Sun, (ESA
  Special Publication, Sheffield), 624

\bibitem[{{Bohlin} \& {Gilliland}(2004)}]{bohlin2004}
{Bohlin}, R.~C. \& {Gilliland}, R.~L. 2004, \aj, 127, 3508

\bibitem[{{Borucki} {et~al.}(2010){Borucki}, {Koch}, {Basri}, {Batalha},
  {Brown}, {Caldwell}, {Caldwell}, {Christensen-Dalsgaard}, {Cochran},
  {DeVore}, {Dunham}, {Dupree}, {Gautier}, {Geary}, {Gilliland}, {Gould},
  {Howell}, {Jenkins}, {Kondo}, {Latham}, {Marcy}, {Meibom}, {Kjeldsen},
  {Lissauer}, {Monet}, {Morrison}, {Sasselov}, {Tarter}, {Boss}, {Brownlee},
  {Owen}, {Buzasi}, {Charbonneau}, {Doyle}, {Fortney}, {Ford}, {Holman},
  {Seager}, {Steffen}, {Welsh}, {Rowe}, {Anderson}, {Buchhave}, {Ciardi},
  {Walkowicz}, {Sherry}, {Horch}, {Isaacson}, {Everett}, {Fischer}, {Torres},
  {Johnson}, {Endl}, {MacQueen}, {Bryson}, {Dotson}, {Haas}, {Kolodziejczak},
  {Van Cleve}, {Chandrasekaran}, {Twicken}, {Quintana}, {Clarke}, {Allen},
  {Li}, {Wu}, {Tenenbaum}, {Verner}, {Bruhweiler}, {Barnes}, \&
  {Prsa}}]{kepler}
{Borucki}, W.~J., {Koch}, D., {Basri}, G., {et~al.} 2010, Science, 327, 977

\bibitem[{{Briquet} {et~al.}(2001){Briquet}, {De Cat}, {Aerts}, \&
  {Scuflaire}}]{briquet2001}
{Briquet}, M., {De Cat}, P., {Aerts}, C., \& {Scuflaire}, R. 2001, \aap, 380,
  177

\bibitem[{{Briquet} {et~al.}(2007){Briquet}, {Hubrig}, {De Cat}, {Aerts},
  {North}, \& {Sch{\"o}ller}}]{briquet2007}
{Briquet}, M., {Hubrig}, S., {De Cat}, P., {et~al.} 2007, \aap, 466, 269

\bibitem[{{Briquet} {et~al.}(2010){Briquet}, {Korhonen}, {Gonz{\'a}lez},
  {Hubrig}, \& {Hackman}}]{briquet2010b}
{Briquet}, M., {Korhonen}, H., {Gonz{\'a}lez}, J.~F., {Hubrig}, S., \&
  {Hackman}, T. 2010, \aap, 511, A71

\bibitem[{{Cantiello} {et~al.}(2010){Cantiello}, {Braithwaite}, {Brandenburg},
  {Del Sordo}, {K{\"a}pyl{\"a}}, \& {Langer}}]{cantiello2010}
{Cantiello}, M., {Braithwaite}, J., {Brandenburg}, A., {et~al.} 2010, ArXiv
  e-prints, 1010.2498

\bibitem[{{Chiar} \& {Tielens}(2006)}]{chiar2006}
{Chiar}, J.~E. \& {Tielens}, A.~G.~G.~M. 2006, \apj, 637, 774

\bibitem[{{Clarke}(2003)}]{clarke2003}
{Clarke}, D. 2003, \aap, 407, 1029

\bibitem[{{Clement}(1969)}]{clement1969}
{Clement}, M.~J. 1969, \apj, 156, 1051

\bibitem[{{Cramer}(1984)}]{cramer1984}
{Cramer}, N. 1984, \aap, 141, 215

\bibitem[{{Cranmer} \& {Owocki}(1995)}]{cranmer1995}
{Cranmer}, S.~R. \& {Owocki}, S.~P. 1995, \apj, 440, 308

\bibitem[{{Cutri} {et~al.}(2003){Cutri}, {Skrutskie}, {van Dyk}, {Beichman},
  {Carpenter}, {Chester}, {Cambresy}, {Evans}, {Fowler}, {Gizis}, {Howard},
  {Huchra}, {Jarrett}, {Kopan}, {Kirkpatrick}, {Light}, {Marsh}, {McCallon},
  {Schneider}, {Stiening}, {Sykes}, {Weinberg}, {Wheaton}, {Wheelock}, \&
  {Zacarias}}]{cat_2mass}
{Cutri}, R.~M., {Skrutskie}, M.~F., {van Dyk}, S., {et~al.} 2003, {2MASS All
  Sky Catalog of point sources, (NASA/IPAC Infrared Science Archive)}

\bibitem[{{De Cat} \& {Aerts}(2002)}]{decat2002}
{De Cat}, P. \& {Aerts}, C. 2002, \aap, 393, 965

\bibitem[{{De Cat} {et~al.}(2005){De Cat}, {Briquet},
  {Daszy{\'n}ska-Daszkiewicz}, {Dupret}, {De Ridder}, {Scuflaire}, \&
  {Aerts}}]{decat2005}
{De Cat}, P., {Briquet}, M., {Daszy{\'n}ska-Daszkiewicz}, J., {et~al.} 2005,
  \aap, 432, 1013

\bibitem[{{Degroote} {et~al.}(2010){Degroote}, {Briquet}, {Auvergne},
  {Simon-Diaz}, {Aerts}, {Noels}, {Rainer}, {Hareter}, {Poretti}, {Mahy},
  {Oreiro}, {Vuckovic}, {Smolders}, {Baglin}, {Baudin}, {Catala}, {Michel}, \&
  {Samadi}}]{degroote2010b}
{Degroote}, P., {Briquet}, M., {Auvergne}, M., {et~al.} 2010, \aap, 519, A38+

\bibitem[{{Droege} {et~al.}(2006){Droege}, {Richmond}, {Sallman}, \&
  {Creager}}]{cat_tass}
{Droege}, T.~F., {Richmond}, M.~W., {Sallman}, M.~P., \& {Creager}, R.~P. 2006,
  \pasp, 118, 1666

\bibitem[{{Dziembowski} \& {Pamiatnykh}(1993)}]{dziembowski1993}
{Dziembowski}, W.~A. \& {Pamiatnykh}, A.~A. 1993, \mnras, 262, 204

\bibitem[{{Gruber} {et~al.}(2009){Gruber}, {Kuschnig}, {Gruberbauer},
  {Hareter}, {Weiss}, \& {Matthews}}]{gruber2009}
{Gruber}, D., {Kuschnig}, R., {Gruberbauer}, M., {et~al.} 2009, Communications
  in Asteroseismology, 158, 217

\bibitem[{{Hauck} \& {Mermilliod}(1998)}]{cat_uvby}
{Hauck}, B. \& {Mermilliod}, M. 1998, \aaps, 129, 431

\bibitem[{{H{\o}g} {et~al.}(2000){H{\o}g}, {Fabricius}, {Makarov}, {Urban},
  {Corbin}, {Wycoff}, {Bastian}, {Schwekendiek}, \& {Wicenec}}]{cat_tycho}
{H{\o}g}, E., {Fabricius}, C., {Makarov}, V.~V., {et~al.} 2000, \aap, 355, L27

\bibitem[{{Hubrig} {et~al.}(2009){Hubrig}, {Briquet}, {De Cat}, {Sch{\"o}ller},
  {Morel}, \& {Ilyin}}]{hubrig2009}
{Hubrig}, S., {Briquet}, M., {De Cat}, P., {et~al.} 2009, Astronomische
  Nachrichten, 330, 317

\bibitem[{{Hubrig} \& {Castelli}(2001)}]{hubrig2001}
{Hubrig}, S. \& {Castelli}, F. 2001, \aap, 375, 963

\bibitem[{{Hussain}(2002)}]{hussain2002}
{Hussain}, G.~A.~J. 2002, Astronomische Nachrichten, 323, 349

\bibitem[{{I{\c s}ik} {et~al.}(2007){I{\c s}ik}, {Sch{\"u}ssler}, \&
  {Solanki}}]{isik2007}
{I{\c s}ik}, E., {Sch{\"u}ssler}, M., \& {Solanki}, S.~K. 2007, \aap, 464, 1049

\bibitem[{{Kurucz}(1993)}]{kurucz1993}
{Kurucz}, R.~L. 1993, VizieR Online Data Catalog, 6039, 0

\bibitem[{{Lanza} {et~al.}(2003){Lanza}, {Rodon{\`o}}, {Pagano}, {Barge}, \&
  {Llebaria}}]{lanza2003}
{Lanza}, A.~F., {Rodon{\`o}}, M., {Pagano}, I., {Barge}, P., \& {Llebaria}, A.
  2003, \aap, 403, 1135

\bibitem[{{Lanza} {et~al.}(1993){Lanza}, {Rodono}, \& {Zappala}}]{lanza1993}
{Lanza}, A.~F., {Rodono}, M., \& {Zappala}, R.~A. 1993, \aap, 269, 351

\bibitem[{{Lefever} {et~al.}(2010){Lefever}, {Puls}, {Morel}, {Aerts}, {Decin},
  \& {Briquet}}]{lefever2010}
{Lefever}, K., {Puls}, J., {Morel}, T., {et~al.} 2010, \aap, 515, A74+

\bibitem[{{Luna} {et~al.}(2008){Luna}, {Cox}, {Satorre}, {Garc{\'{\i}}a
  Hern{\'a}ndez}, {Su{\'a}rez}, \& {Garc{\'{\i}}a Lario}}]{luna2008}
{Luna}, R., {Cox}, N.~L.~J., {Satorre}, M.~A., {et~al.} 2008, \aap, 480, 133

\bibitem[{{Ma{\'{\i}}z-Apell{\'a}niz}(2007)}]{maizapellaniz2007}
{Ma{\'{\i}}z-Apell{\'a}niz}, J. 2007, in The Future of Photometric,
  Spectrophotometric and Polarimetric Standardization, ed. {C.~Sterken (San
  Fransisco, ASP)}, 364, 227

\bibitem[{{Mermilliod} {et~al.}(1997){Mermilliod}, {Mermilliod}, \&
  {Hauck}}]{cat_gcpd}
{Mermilliod}, J., {Mermilliod}, M., \& {Hauck}, B. 1997, \aaps, 124, 349

\bibitem[{{Miglio} {et~al.}(2007{\natexlab{a}}){Miglio}, {Montalb{\'a}n}, \&
  {Dupret}}]{miglio2007}
{Miglio}, A., {Montalb{\'a}n}, J., \& {Dupret}, M.-A. 2007{\natexlab{a}},
  \mnras, 375, L21

\bibitem[{{Miglio} {et~al.}(2007{\natexlab{b}}){Miglio}, {Montalb{\'a}n}, \&
  {Dupret}}]{miglio2007a}
{Miglio}, A., {Montalb{\'a}n}, J., \& {Dupret}, M.-A. 2007{\natexlab{b}},
  Communications in Asteroseismology, 151, 48

\bibitem[{{Monet} {et~al.}(2003){Monet}, {Levine}, {Canzian}, {Ables}, {Bird},
  {Dahn}, {Guetter}, {Harris}, {Henden}, {Leggett}, {Levison}, {Luginbuhl},
  {Martini}, {Monet}, {Munn}, {Pier}, {Rhodes}, {Riepe}, {Sell}, {Stone},
  {Vrba}, {Walker}, {Westerhout}, {Brucato}, {Reid}, {Schoening}, {Hartley},
  {Read}, \& {Tritton}}]{cat_usnob1}
{Monet}, D.~G., {Levine}, S.~E., {Canzian}, B., {et~al.} 2003, \aj, 125, 984

\bibitem[{{Moon} \& {Dworetsky}(1985)}]{moon1985}
{Moon}, T.~T. \& {Dworetsky}, M.~M. 1985, \mnras, 217, 305

\bibitem[{{Niemczura} {et~al.}(2009){Niemczura}, {Morel}, \&
  {Aerts}}]{niemczura2009}
{Niemczura}, E., {Morel}, T., \& {Aerts}, C. 2009, \aap, 506, 213

\bibitem[{{Ofek}(2008)}]{cat_pasp}
{Ofek}, E.~O. 2008, \pasp, Chicago, 120, 1128

\bibitem[{{Palacios} {et~al.}(2010){Palacios}, {Gebran}, {Josselin}, {Martins},
  {Plez}, {Belmas}, \& {L{\`e}bre}}]{palacios2010}
{Palacios}, A., {Gebran}, M., {Josselin}, E., {et~al.} 2010, \aap, 516, A13+

\bibitem[{{Preibisch} {et~al.}(2001){Preibisch}, {Weigelt}, \&
  {Zinnecker}}]{preibisch2001}
{Preibisch}, T., {Weigelt}, G., \& {Zinnecker}, H. 2001, in The Formation of
  Binary Stars, 200, 69

\bibitem[{{Raskin} {et~al.}(2011){Raskin}, {van Winckel}, {Hensberge},
  {Jorissen}, {Lehmann}, {Waelkens}, {Avila}, {de Cuyper}, {Degroote},
  {Dubosson}, {Dumortier}, {Fr{\'e}mat}, {Laux}, {Michaud}, {Morren}, {Perez
  Padilla}, {Pessemier}, {Prins}, {Smolders}, {van Eck}, \& {Winkler}}]{hermes}
{Raskin}, G., {van Winckel}, H., {Hensberge}, H., {et~al.} 2011, \aap, 526,
  A69+

\bibitem[{{Reese} {et~al.}(2009){Reese}, {MacGregor}, {Jackson}, {Skumanich},
  \& {Metcalfe}}]{reese2009}
{Reese}, D.~R., {MacGregor}, K.~B., {Jackson}, S., {Skumanich}, A., \&
  {Metcalfe}, T.~S. 2009, \aap, 506, 189

\bibitem[{{Reiners} \& {Royer}(2004{\natexlab{a}})}]{reiners2004b}
{Reiners}, A. \& {Royer}, F. 2004{\natexlab{a}}, \aap, 428, 199

\bibitem[{{Reiners} \& {Royer}(2004{\natexlab{b}})}]{reiners2004}
{Reiners}, A. \& {Royer}, F. 2004{\natexlab{b}}, \aap, 415, 325

\bibitem[{{Reiners} \& {Schmitt}(2002{\natexlab{a}})}]{reiners2002b}
{Reiners}, A. \& {Schmitt}, J.~H.~M.~M. 2002{\natexlab{a}}, \aap, 388, 1120

\bibitem[{{Reiners} \& {Schmitt}(2002{\natexlab{b}})}]{reiners2002}
{Reiners}, A. \& {Schmitt}, J.~H.~M.~M. 2002{\natexlab{b}}, \aap, 384, 155

\bibitem[{{Reiners} \& {Schmitt}(2003)}]{reiners2003}
{Reiners}, A. \& {Schmitt}, J.~H.~M.~M. 2003, \aap, 412, 813

\bibitem[{{Royer} {et~al.}(2007){Royer}, {Zorec}, \& {G{\'o}mez}}]{royer2007}
{Royer}, F., {Zorec}, J., \& {G{\'o}mez}, A.~E. 2007, \aap, 463, 671

\bibitem[{{Scuflaire} {et~al.}(2008){Scuflaire}, {Th{\'e}ado}, {Montalb{\'a}n},
  {Miglio}, {Bourge}, {Godart}, {Thoul}, \& {Noels}}]{scuflaire2008}
{Scuflaire}, R., {Th{\'e}ado}, S., {Montalb{\'a}n}, J., {et~al.} 2008, \apss,
  316, 83

\bibitem[{{Shorlin} {et~al.}(2002){Shorlin}, {Wade}, {Donati}, {Landstreet},
  {Petit}, {Sigut}, \& {Strasser}}]{shorlin2002}
{Shorlin}, S.~L.~S., {Wade}, G.~A., {Donati}, J.-F., {et~al.} 2002, \aap, 392,
  637

\bibitem[{{Sim{\'o}n-D{\'{\i}}az} \& {Herrero}(2007)}]{simondiaz2007}
{Sim{\'o}n-D{\'{\i}}az}, S. \& {Herrero}, A. 2007, \aap, 468, 1063

\bibitem[{{Solano} {et~al.}(2005){Solano}, {Catala}, {Garrido}, {Poretti},
  {Janot-Pacheco}, {Guti{\'e}rrez}, {Gonz{\'a}lez}, {Mantegazza}, {Neiner},
  {Fremat}, {Charpinet}, {Weiss}, {Amado}, {Rainer}, {Tsymbal}, {Lyashko},
  {Ballereau}, {Bouret}, {Hua}, {Katz}, {Ligni{\`e}res}, {L{\"u}ftinger},
  {Mittermayer}, {Nesvacil}, {Soubiran}, {van't Veer-Menneret}, {Goupil},
  {Costa}, {Rolland}, {Antonello}, {Bossi}, {Buzzoni}, {Rodrigo}, {Aerts},
  {Butler}, {Guenther}, \& {Hatzes}}]{solano2005}
{Solano}, E., {Catala}, C., {Garrido}, R., {et~al.} 2005, \aj, 129, 547

\bibitem[{{Strassmeier} \& {Bopp}(1992)}]{strassmeier1992}
{Strassmeier}, K.~G. \& {Bopp}, B.~W. 1992, \aap, 259, 183

\bibitem[{{Thompson} {et~al.}(1978){Thompson}, {Nandy}, {Jamar}, {Monfils},
  {Houziaux}, {Carnochan}, \& {Wilson}}]{cat_td1}
{Thompson}, G.~I., {Nandy}, K., {Jamar}, C., {et~al.} 1978, {Catalogue of
  stellar ultraviolet fluxes. A compilation of absolute stellar fluxes measured
  by the Sky Survey Telescope (S2/68) aboard the ESRO satellite TD-1,
  (NASA/STI)}

\bibitem[{{Uytterhoeven} {et~al.}(2005){Uytterhoeven}, {Briquet}, {Aerts},
  {Telting}, {Harmanec}, {Lefever}, \& {Cuypers}}]{uytterhoeven2005}
{Uytterhoeven}, K., {Briquet}, M., {Aerts}, C., {et~al.} 2005, \aap, 432, 955

\bibitem[{{van der Bliek} {et~al.}(1996){van der Bliek}, {Manfroid}, \&
  {Bouchet}}]{cat_eso}
{van der Bliek}, N.~S., {Manfroid}, J., \& {Bouchet}, P. 1996, \aaps, 119, 547

\bibitem[{{von Zeipel}(1924)}]{vonzeipel1924}
{von Zeipel}, H. 1924, \mnras, 84, 665

\bibitem[{{Wade} {et~al.}(2006){Wade}, {Auri{\`e}re}, {Bagnulo}, {Donati},
  {Johnson}, {Landstreet}, {Ligni{\`e}res}, {Marsden}, {Monin}, {Mouillet},
  {Paletou}, {Petit}, {Toqu{\'e}}, {Alecian}, \& {Folsom}}]{wade2006}
{Wade}, G.~A., {Auri{\`e}re}, M., {Bagnulo}, S., {et~al.} 2006, \aap, 451, 293

\bibitem[{{Waelkens}(1991)}]{waelkens1991}
{Waelkens}, C. 1991, \aap, 246, 453

\bibitem[{{Wolff}(1983)}]{wolff1983}
{Wolff}, S.~C. 1983, {The A-stars: Problems and perspectives. Monograph series
  on nonthermal phenomena in stellar atmospheres, (NASA/STI)}

\bibitem[{{Zorec} {et~al.}(2011){Zorec}, {Fr{\'e}mat}, {Domiciano de Souza},
  {Delaa}, {Stee}, {Mourard}, {Cidale}, {Martayan}, {Georgy}, \&
  {Ekstr{\"o}m}}]{zorec2011}
{Zorec}, J., {Fr{\'e}mat}, Y., {Domiciano de Souza}, A., {et~al.} 2011, \aap,
  526, A87+

\end{thebibliography}

\Online
\begin{appendix}
 \section{Light curves and frequency spectra of B8/B9 stars observed with CoRoT}

\begin{figure*} 
\includegraphics[width=2\columnwidth]{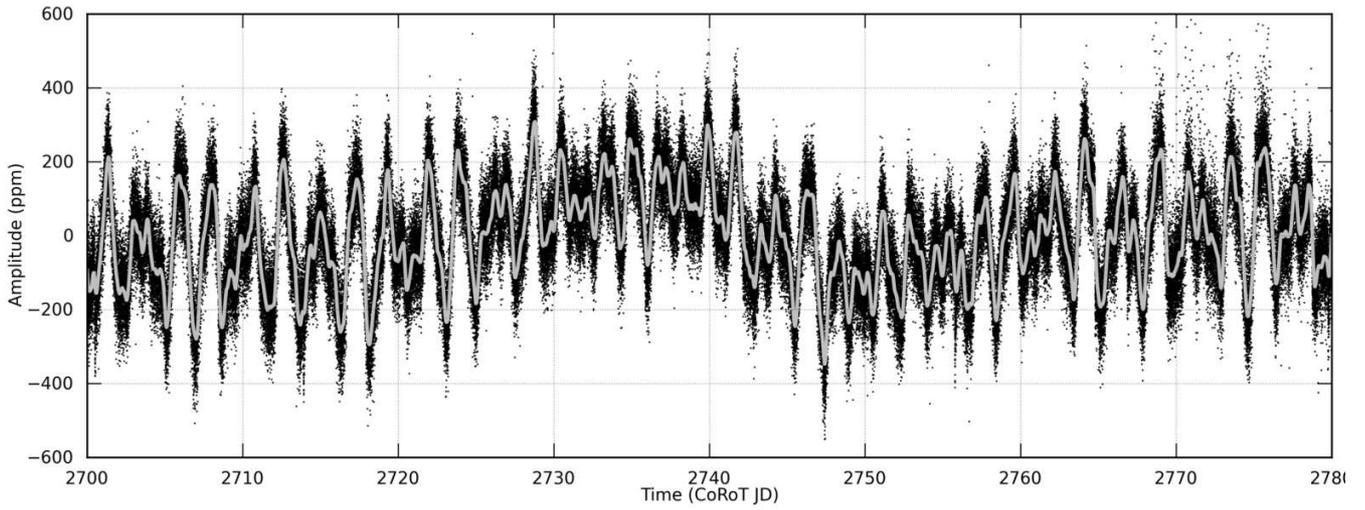} 
 \caption{Excerpt of the CoRoT light curve of HD\,181440, showing multiperiodic variability (grey line shows a fit to the light curve, black dots represent to CoRoT data).}
\label{fig:hd181440} 
\end{figure*} 

\begin{figure*} 
\includegraphics[width=2\columnwidth]{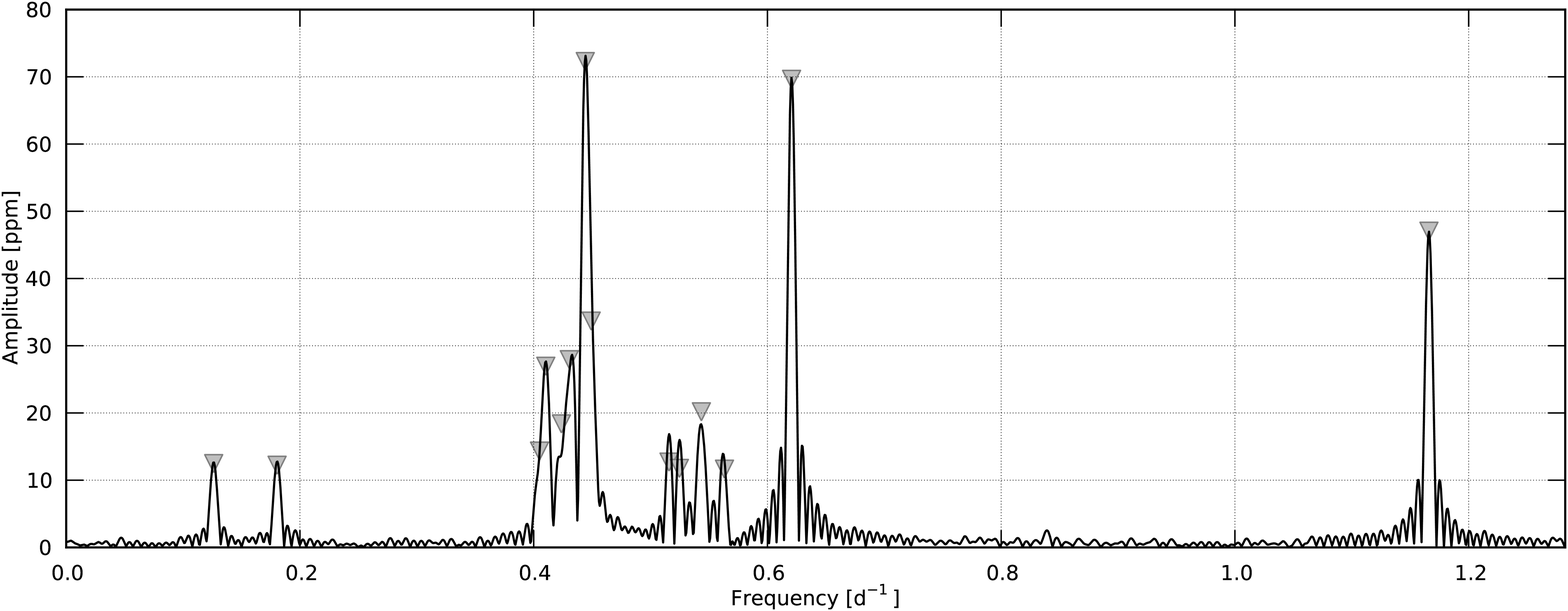} 
 \caption{Frequency spectrum of the CoRoT light curve of HD\,181440 above 0.1\,d$^{-1}$ (black). The significant frequencies and fitted amplitudes above 10 ppm are indicated in grey.}
\label{fig:per:hd181440} 
\end{figure*} 

\begin{figure*} 
\includegraphics[width=2\columnwidth]{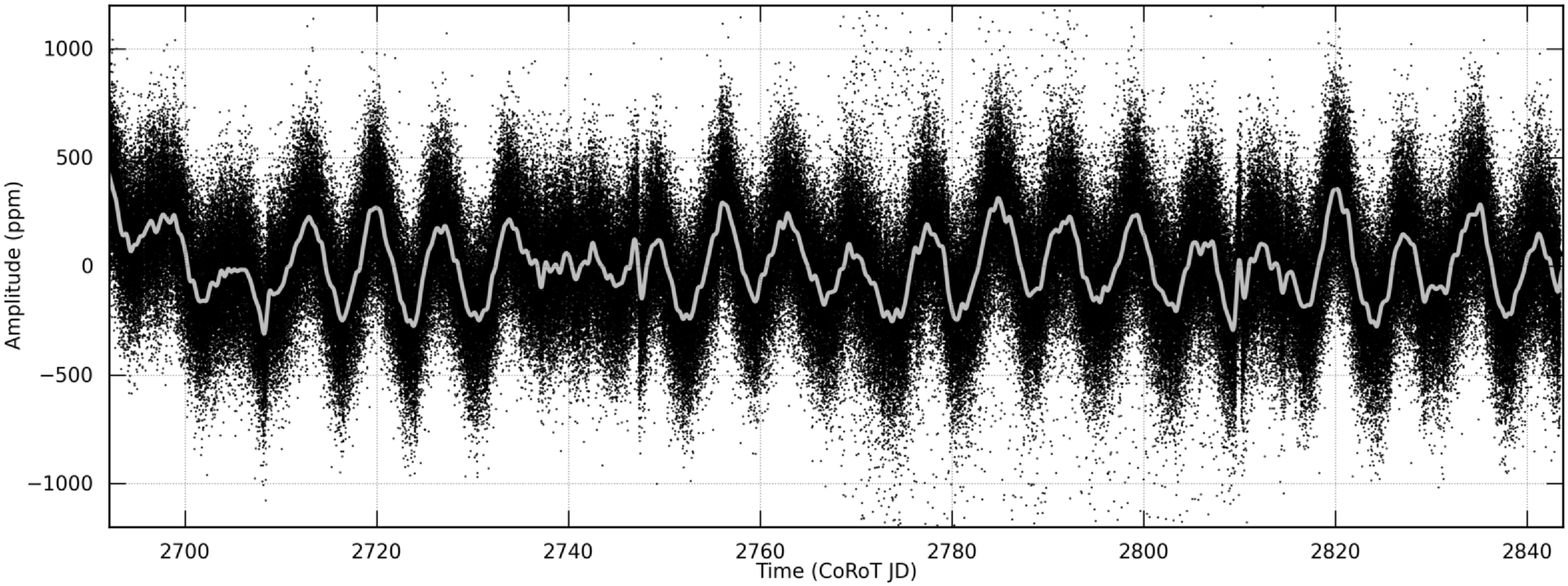} 
 \caption{Excerpt of the CoRoT light curve of HD\,182198, with only one clear clear frequency (grey line shows a fit to the light curve, black dots represent to CoRoT data).}\label{fig:hd182198} 
\end{figure*}

 \begin{figure*} 
\includegraphics[width=2\columnwidth]{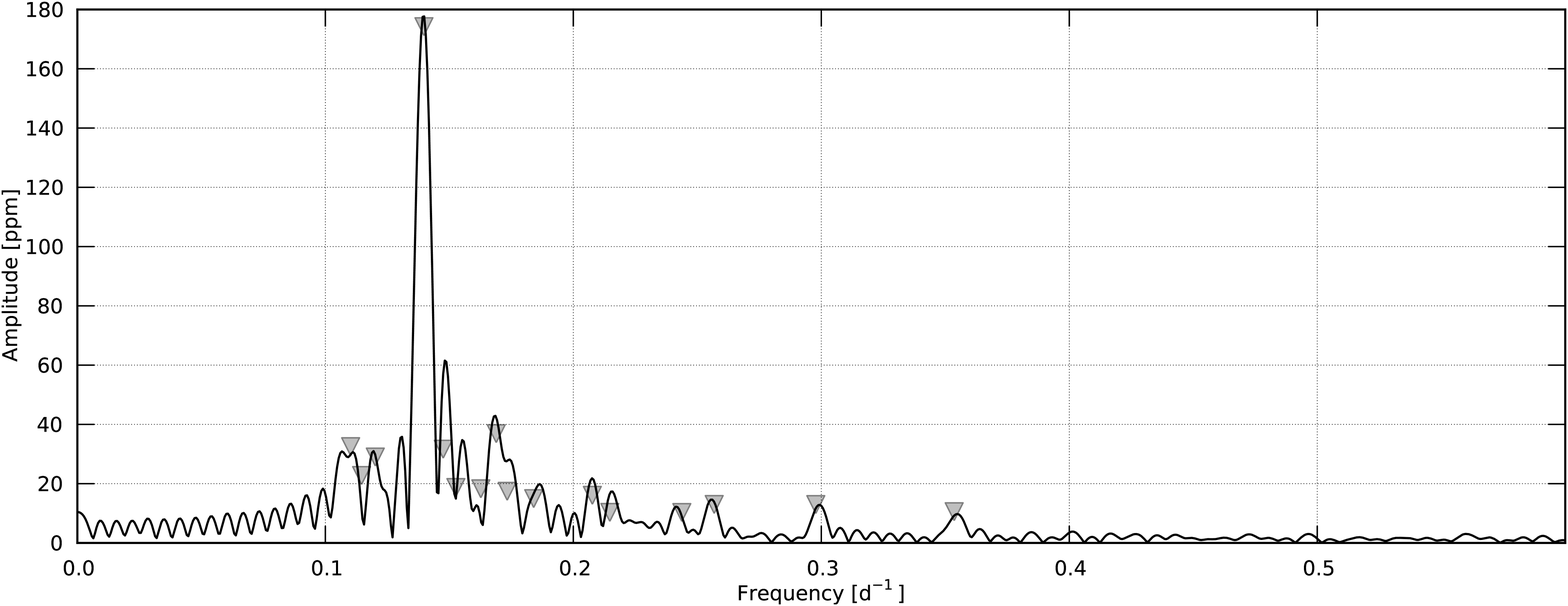} 
 \caption{Frequency spectrum of the CoRoT light curve of HD\,182198 above 0.1\,d$^{-1}$ (black). The significant frequencies and fitted amplitudes above 10 ppm are indicated in grey.}
\label{fig:per:hd182198} 
\end{figure*}

\begin{figure*} 
\includegraphics[width=2\columnwidth]{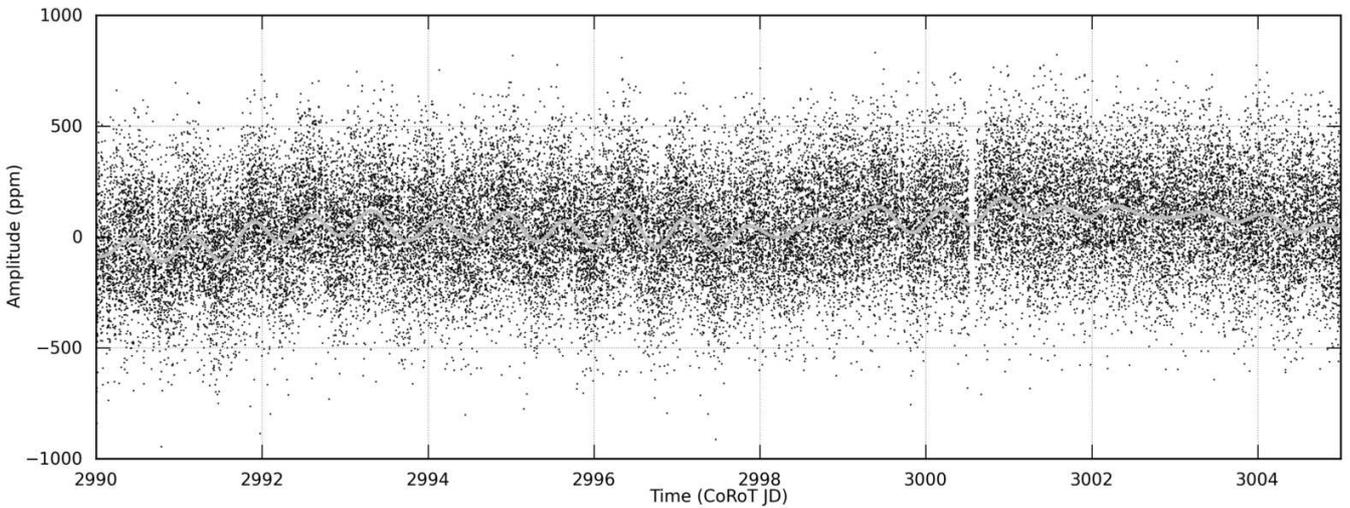} 
 \caption{Excerpt of the CoRoT light curve of HD\,49677, with only one clear frequency detected (grey line shows a fit to the light curve, black dots represent to CoRoT data).}
\label{fig:hd149677} 
\end{figure*}

\begin{figure*} 
\includegraphics[width=2\columnwidth]{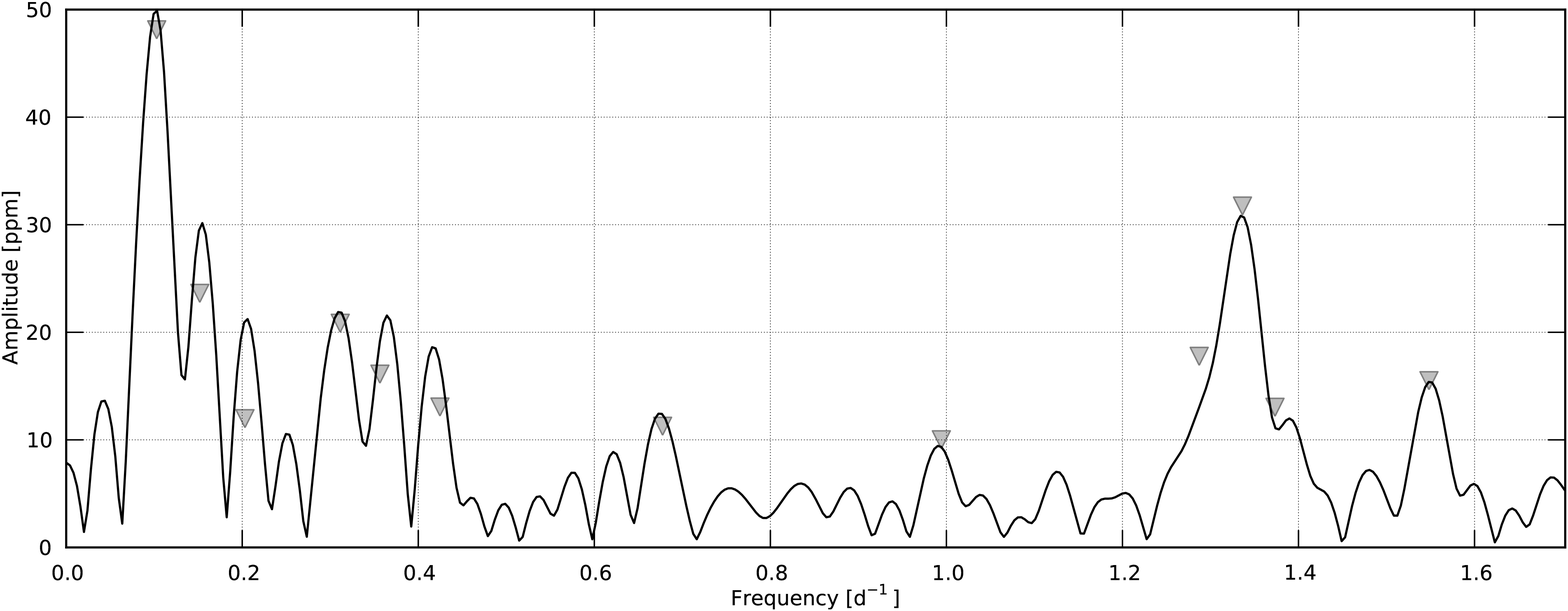} 
 \caption{Frequency spectrum of the CoRoT light curve of HD\,49677 above 0.1\,d$^{-1}$ (black). The significant frequencies with amplitudes above 10 ppm are indicated in grey.}
\label{fig:per:hd49677} 
\end{figure*}  

\begin{figure*} 
\includegraphics[width=2\columnwidth]{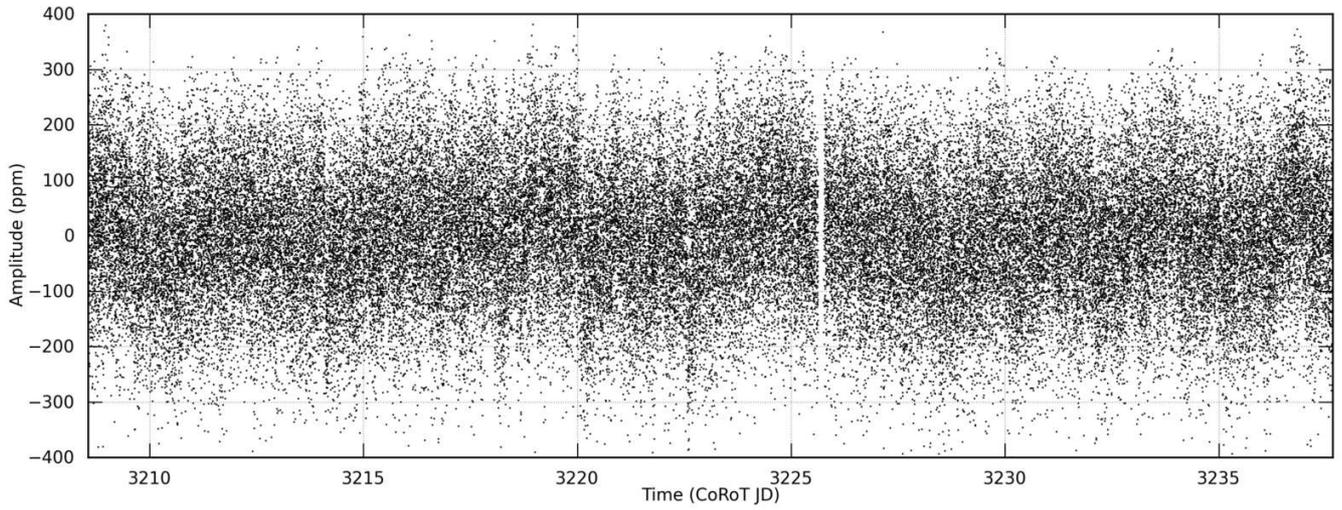} 
 \caption{Excerpt of the CoRoT light curve of HD\,46179. No clear signs of variability are detected (grey line shows a fit to the light curve, black dots represent to CoRoT data).}
\label{fig:hd46179} 
\end{figure*} 

\begin{figure*} 
\includegraphics[width=2\columnwidth]{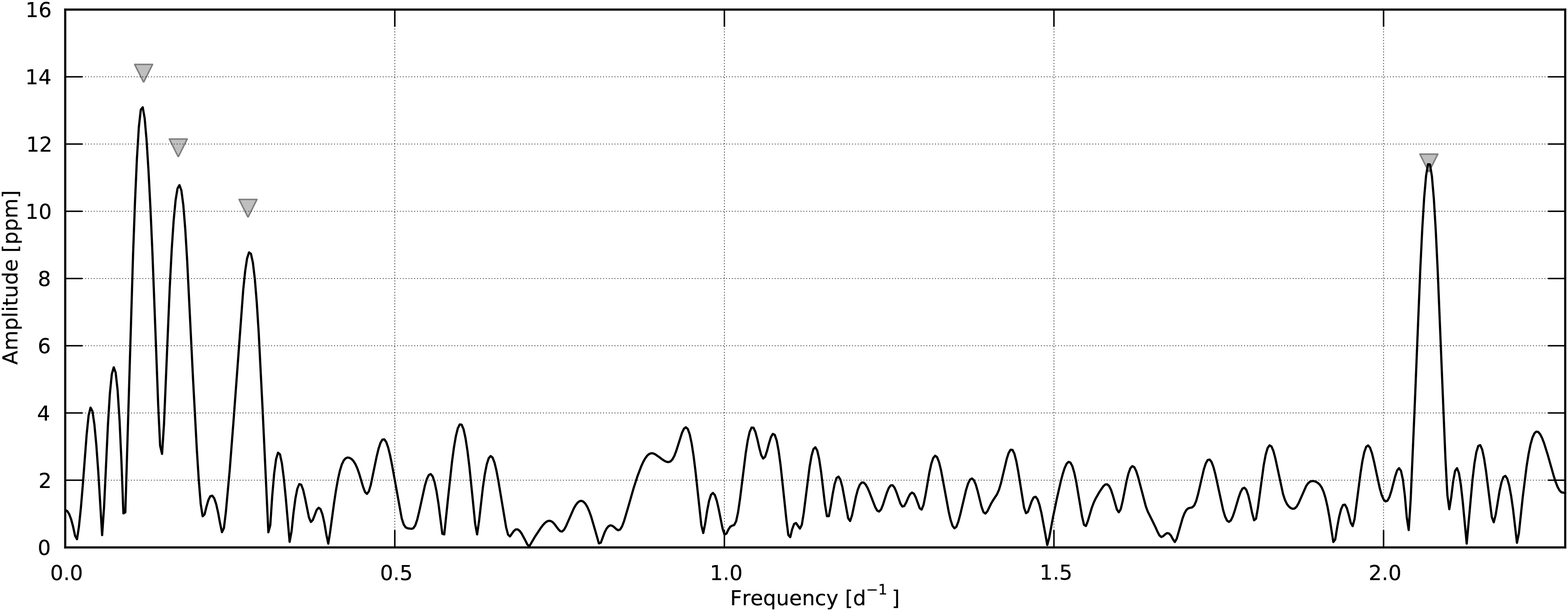}
 \caption{Frequency spectrum of the CoRoT light curve of HD\,46179 above 0.1\,d$^{-1}$ (black). The significant frequencies with amplitudes above 10 ppm are indicated in grey.}
\label{fig:per:hd46179} 
\end{figure*}

\end{appendix}

\end{document}